\documentclass[preprint,5p,twocolumn,11pt,sort&compress]{elsarticle}
\usepackage[english]{babel}
\usepackage{graphicx}
\usepackage[usenames,dvipsnames,svgnames,table]{xcolor}
\usepackage{commath}
\usepackage{bm}
\usepackage{amsfonts}
\usepackage{xcolor}
\usepackage[colorlinks = true,
            linkcolor = blue,
            urlcolor  = blue,
            citecolor = blue,
            anchorcolor = blue]{hyperref}

\def\bea{\begin{eqnarray}}
\def\eea{\end{eqnarray}}
\def\fsig{f\sigma_8}
\def\sig{\sigma_8}

\def\ff{\mathfrak{f}}
\def\d{{\rm d}}

\makeatletter
\def\ps@pprintTitle{%
  \let\@oddhead\@empty
  \let\@evenhead\@empty
  \let\@oddfoot\@empty
  \let\@evenfoot\@oddfoot
}
\makeatother

\begin{document}

\title{Multi-tasking the growth of cosmological structures}
\date{}

\author[label1]{Louis Perenon \corref{cor}}
\cortext[cor]{Corresponding author: perenon.louis@yahoo.fr}
\address[label1]{Department of Physics \& Astronomy, University of the Western Cape, Cape Town 7535, South Africa}

\author[label2]{Matteo Martinelli}
\address[label2]{Instituto de F\'isica Te\'orica UAM-CSIC, Campus de Cantoblanco, E-28049 Madrid, Spain}

\author[label3,label4]{St\'ephane Ili\'c}
\address[label3]{Universit\'e PSL, Observatoire de Paris, Sorbonne Universit\'e, CNRS, LERMA, F-75014, Paris, France}

\address[label4]{Université de Paris, CNRS, Astroparticule et Cosmologie, F-75013 Paris, France}

\author[label1,label5]{Roy Maartens}
\address[label5]{Institute of Cosmology \& Gravitation, University of Portsmouth, Portsmouth PO1 3FX, UK}

\author[label1,label6]{Michelle Lochner}
\address[label6]{South African Radio Astronomy Observatory, Cape Town 7925, South Africa}

\author[label7,label1]{Chris Clarkson}
\address[label7]{School of Physics \& Astronomy, Queen Mary University of London, London E1 4NS, UK}

\begin{abstract}
Next-generation large-scale structure surveys will deliver a significant increase in the precision of growth data, allowing us to use `agnostic' methods to study the evolution of perturbations without the assumption of a cosmological model. We focus on a particular machine learning tool, Gaussian processes, to reconstruct the growth rate $f$, the root mean square of matter fluctuations $\sigma_8$, and their product $f\sigma_8$. We apply this method to simulated data, representing the precision of upcoming Stage IV galaxy surveys. We extend the standard single-task approach to a multi-task approach that reconstructs the three functions simultaneously, thereby taking into account their inter-dependence. We find that this multi-task approach outperforms the precision of the single-task approach for future surveys. By contrast, the limited sensitivity of current data severely hinders the use of agnostic methods, since the Gaussian processes parameters need to be fine tuned in order to obtain robust reconstructions.
\end{abstract}


\maketitle
\tableofcontents

\section{Introduction}

Galaxy surveys have provided observations of  the growth of large-scale structure  with steadily increasing precision and range. This trend is set  to accelerate in the coming years with next-generation spectroscopic surveys such as those with {\em Euclid} \cite{Blanchard:2019oqi}, the Square Kilometre Array (SKA) \cite{Bacon:2018dui}, the Dark Energy Spectroscopic Instrument (DESI) \cite{Levi:2019ggs}, and the Nancy Grace Roman Space Telescope \cite{Dore:2019pld}. The evolution of matter perturbations gives direct insight into the underlying theory of gravity. Whether the correct description is  the standard model of cosmology $\Lambda$CDM, a dynamical dark energy model or modified gravity  (see e.g. \cite{Capozziello:2011et,Clifton:2011jh} for reviews), is still an open question.   

Studying gravity on cosmological scales has been pursued via two paths in recent years. The model-dependent approach constrains a particular model of dark energy or modified gravity with observations and compares it to $\Lambda$CDM. This approach has shown hints that alternative theories may be favoured via Bayesian evidence considerations \cite{Peirone:2019aua,Sola:2019jek,Aoki:2020oqc,DiValentino:2020kpf}. The `agnostic' approach avoids choosing a particular model and develops predictions as model independent as possible, which can be used to evaluate how they differ from those of a given model. 

Machine learning offers several model-agnostic methods, including Gaussian Process (GP) regressions, which enable data-driven reconstructions of trends \cite{Rasmussen}. The characteristic input of GP is statistical and corresponds to a kernel to build a covariance matrix and a mean prior. The use of GP is now common in cosmology and has been focused mainly on the background evolution of the universe \cite{Holsclaw:2010nb,Holsclaw:2010sk,Holsclaw:2011wi,Shafieloo:2012ht,Seikel:2012uu,Yahya:2013xma,Bester:2013fya,Busti:2014dua,Bester:2015gla,Bester:2016fbs,Joudaki:2017zhq,Gomez-Valent:2018hwc,Haridasu:2018gqm,Gerardi:2019obr,Keeley:2019hmw,Bengaly:2019oxx,Bengaly:2019ibu,Aljaf:2020eqh,Keeley:2019hmw,Colgain:2021ngq,Dhawan:2021mel,Mukherjee:2021kcu,2021arXiv210407356V,Canas-Herrera:2021qxs}. GP has also proved useful to determine the value of the Hubble constant using strong lensing data and supernova data \cite{Liao:2019qoc,Liu:2019ddm,Pandey:2019yic,Liao:2020zko,Renzi:2020fnx,Bengaly:2020neu,Zhang:2021tmg}, to improve photometric redshift estimates \cite{2016MNRAS.462..726A}, and to speed up the processing of weak lensing data \cite{Mootoovaloo:2020ott}. Less work has been done with GP on measurements of perturbations. Examples are reconstructions of the linear anisotropic stress  \cite{Pinho:2018unz} and of the growth rate of large-scale structure \cite{Zhang:2018gjb,Yin:2018mvu,Li:2019nux,Benisty:2020kdt}. 

In this paper we use GP to reconstruct the growth rate of large-scale structure. We produce forecasts for a nominal Stage IV spectroscopic survey. Our focus is not only on the growth variable $\fsig$, but also on the separated growth rate $f$ and variance of matter fluctuations $\sig$. We apply the concept of multi-task GP \cite{Caruana1998,Rasmussen,866abfcac5f9481d97628a546255878b,21ae58e5ad934d7494d64e5bfd4a1a52,Melkumyan,Vasudevan}, previously used in cosmology to improve reconstructions of the background acceleration \cite{Haridasu:2018gqm}.

The paper is structured as follows. We review the basics of the growth of large-scale structure and the methodology of our analysis in \autoref{sec:context}. We then analyse and compare the single- and multi-task reconstructions of mock data in \autoref{sec:perfo}, where we display the improvement from using multi-task GP. Previous single-task GP studies of growth rate in \cite{Zhang:2018gjb,Yin:2018mvu,Li:2019nux,Benisty:2020kdt} do not give much detail about the assumptions specific to their reconstruction of $\fsig$. In \autoref{sec:current}, we discuss difficulties in reproducing such results and how we find current growth data to be ill-suited for robust GP reconstructions. We conclude in \autoref{sec:discussion}.

\section{Context}\label{sec:context}

We briefly review the growth of large-scale structure before presenting the GP reconstructions from the single-task to multi-task approach. Then we show how our mock data are derived and the numerical tools that we employ.

\subsection{Growth of large-scale structure}\label{sec:growthgen}

The galaxy distribution in redshift space is distorted  by galaxy peculiar velocities that are induced by the growth of cosmological structures, an effect known as redshift-space distortions (RSD) \cite{1977ApJ...212L...3S,Kaiser:1987qv,Guzzo:2008ac,Song:2008qt,Percival:2008sh}. RSD measurements determine the product of  the growth rate $f$ and variance of matter fluctuations $\sigma_8^2$, where
\bea
\label{eq:defD}
    f(z)&=&-\frac{\d\ln{D(z)}}{\d\ln(1+z)}\, ,\\
\sigma^2_8(z) &=&\frac{1}{2\pi^2}\int_0^{\infty} P_m(k,z) \, \hat{W}^2(k)\, k^2 \, \d k\, .
\eea 
Here $D(z)=\delta_m (\bm k,z)/\delta_m(\bm k,0)$ is the linear growth factor of the matter density contrast $\delta_m$, $P_m$ is the matter power spectrum and $\hat{W}$ is the Fourier transform of the top-hat window function that  smooths over spheres of radius $8\,{\rm Mpc}/h$.

Both these quantities depend on the evolution of $\delta_m$, whose redshift dependence and sometimes scale dependence  will differ between models of gravity. The reconstruction of $f$ and $\sigma_8$ is then an ideal tool to distinguish between  different theories. In general relativity with standard dark energy, the linear growth rate does not depend on the scale $k$, but this does not necessarily hold in alternative theories of gravity.

RSD are not sensitive to the two functions $f$ and $\sigma_8$ separately, but to their product $\fsig$. Disentangling $f$ and $\sigma_8$ given a measurement of $\fsig$, requires their degeneracy to be broken.  It is possible to break this degeneracy through two different techniques, both requiring additional observations: (i)~by combining RSD measurements in the power spectrum and bispectrum \cite{Gil-Marin:2016wya}; (ii)~by combining RSD data with galaxy-galaxy lensing data \cite{delaTorre:2016rxm,Shi:2017qpr,Jullo:2019lgq}. 

The use of GP to reconstruct trends from current $\fsig$ data has been studied previously in \cite{Zhang:2018gjb,Yin:2018mvu,Li:2019nux,Benisty:2020kdt}. Given the low precision and relative randomness of current data, we find that the reconstructions depend too strongly on the GP assumptions made. We expand on this  in \autoref{sec:current}. 

\subsection{Basics of Gaussian processes}\label{sec:stgp}

Reconstructing a latent function from data with GP essentially corresponds to assuming that the data is a random realisation of the function with Gaussian noise given by the covariance matrix of the data. The reconstruction of $\bm \ff^*$ as a function of $\bm X^*$, given data {$\bm y$} as a function of  $\bm X$, with a covariance matrix $C$, is characterised by the joint probability distribution \cite{Rasmussen}
\begin{equation}\label{eq:jointdistr}
\begin{bmatrix} \bm y \\ \bm \ff^* \end{bmatrix}\! \sim \!
\mathcal{N}\! \left(\! 
\begin{bmatrix}
\bm \mu  \\
\bm \mu^*
\end{bmatrix}
\!,
\begin{bmatrix}
K(\bm X,\bm X) +C & K(\bm X,\bm X^*)\\
K(\bm X^*,\bm X)    & K(\bm X^*,\bm X^*)
\end{bmatrix}\!
\right)\!. 
\end{equation}
Here $\bm \mu$ and $\bm \mu^*$ correspond to  priors on the means of the reconstructed functions  of $\bm X^*$ and $\bm X$ respectively. $K$ is the kernel of the GP, i.e. the covariance function that relates the values of the function to be reconstructed over the points $x$ in $\bm X$ and $x^*$ in $\bm X^*$. From the joint distribution of \autoref{eq:jointdistr}, the mean and covariance of the reconstructed function are 
\begin{align}\label{eq:gpmean}
{\rm mean}(\ff^*)  &=  \bm \mu^* \\\nonumber
& + K(\bm X^*, \bm X)\!\left[\!K(\bm X,\bm X) + C\right]^{-1}\!\! (\bm y -\bm \mu),  \\[2mm]
\label{eq:gpcov}
{\rm cov}(\ff^*)  & = K(\bm X^*,\bm X^*) \\\nonumber
& - K(\bm X^*,\bm X)\! \left[\!K(\bm X,\bm X) + C\right]^{-1} K(\bm X,\bm X^*) .
\end{align}
The marginal log likelihood of the reconstructed function can then be obtained as
\begin{align}\label{eq:gp_log_marginal}
\ln \mathcal{L} = & -\dfrac{1}{2}(\bm y - \bm \mu)^T \left[\! K(\bm X,\bm X) + C \right]^{-1} (\bm y - \bm \mu) \nonumber\\
& - \dfrac{1}{2}\ln\left| K(\bm X,\bm X) + C \right| - \frac{n}{2}\ln 2\pi  \;,
\end{align}
where $n$ is the number of data points. 

It should  now be clear that a GP allows us, given a set of training data points, to reconstruct a cosmological function without the need to assume its trend or a parametrisation, but only relying on the assumption that the data are a Gaussian realisation of the underlying function. For this reason, GP reconstructions fall in the category of non-parametric methods.  However,  this does not imply that GP are independent of  free quantities to be fixed. A kernel must be chosen, which depends on a set of `hyperparameters' that need to be evaluated during the reconstruction process. In addition, the reconstruction also requires a mean prior function.

The kernel most often used in cosmology so far is the squared-exponential
\begin{align}\label{eq:SE}
k(x,x') & = \sigma^2 \exp\Big(-\dfrac{\vert x-x' \vert ^{2}}{2 \xi^{2}}\Big).
\end{align}
Two hyperparameters are related to the characteristic scales in the data:  the correlation length  $\xi$ and signal variance $\sigma$ \cite{Seikel:2013fda}. (See Chapter 4 of \cite{Rasmussen} for more details on covariance functions.)

\subsection{Multi-tasking the growth of structures}\label{sec:mtgp}

The previous section describes how to reconstruct one function from a data set, i.e. a single-task GP. In order to generalise this for several data sets and latent functions which describe the same underlying phenomena, we use a multi-task GP \cite{Caruana1998,Rasmussen}. This extension  properly accommodates possible correlations between the data sets for each function. As described in \autoref{sec:growthgen}, it is straightforward to obtain $\fsig$ measurements, while a more involved procedure is then needed to obtain the disentangled $f$ and $\sig$ contributions. If these data sets originate from different surveys and different methods, the data would in principle not be correlated between components. In general there will be correlations between the data sets. The multi-task GP is able to capture such correlations.

In addition to a joint likelihood that incorporates correlations between the three data sets, multi-task GP also allows us to takes into account that the growth functions are inter-dependent, since all three are probes of the evolution of matter perturbations. Furthermore, one function is a product of the other two.

In the general case, a combined data set including $f$, $\sig$ and $\fsig$ data has a covariance matrix in block form,
\begin{equation}
C=\!
  \left[\! \begin{array}{lll}
   {{\rm cov}(f,f)}     & {\rm cov}(f,\sig)     & {\rm cov}(f,\fsig)    \\
   {\rm cov}(f,\sig)  & {{\rm cov}(\sig,\sig)}  & {\rm cov}(\sig,\fsig) \\
   {\rm cov}(f,\fsig) & {\rm cov}(\sig,\fsig) & {{\rm cov}(\fsig,\fsig)}     
  \end{array} \! \right],
\end{equation}
where diagonal entries are  auto-covariances  and off-diagonal are cross-covariances.  Similarly, the multi-task GP covariance function encodes not only the three auto-covariance functions, but also the cross-covariances arising from their  inter-dependence as theoretical probes of growth. It has the form
 \begin{equation}\label{eq:Ktilde}
\tilde{K} = 
  \left[ {\begin{array}{lll}
   K_{f,f}      & K_{f,\sig}    & K_{f,\fsig}     \\
   K_{f,\sig}  & K_{\sig,\sig}  & K_{\sig,\fsig}  \\
   K_{f,\fsig} & K_{\sig,\fsig} & K_{\fsig,\fsig}       \\
  \end{array} } \right] + C .
\end{equation}
Provided the data covariance matrix $C$ is block diagonal, setting all non-diagonal blocks of $\tilde{K}$ to zero leads back to the standard approach where each latent function for each data set would be reconstructed as a `single-task', i.e. independently of the other sets. In practice, this is indeed equivalent to considering separately each single-task $i$, where the relevant covariance matrix for the GP is $K_i+C_i$, and doing the respective reconstructions. By contrast, writing the full $\tilde{K}$ as in \autoref{eq:Ktilde} allows multiple functions to be reconstructed simultaneously, including the covariance between the different tasks \cite{Haridasu:2018gqm,866abfcac5f9481d97628a546255878b,21ae58e5ad934d7494d64e5bfd4a1a52,Melkumyan,Vasudevan}.

The $\tilde{K}$ kernel of the multi-task GP is constructed as follows. First, the sub-kernels for each diagonal  latent function must be chosen. Then, the corresponding cross-kernels are obtained from the convolution of the basis functions of each  sub-kernel (see for instance \cite{Melkumyan} and Section 2.2 of \cite{Haridasu:2018gqm} for details). The multi-task GP construction implies that the final reconstruction is characterised by a set of hyperparameters brought by the kernels describing the auto-covariance of each latent function, as in the single-task approach, while the cross-terms do not contribute any additional parameter to the reconstruction. For the case of the squared exponential kernel \autoref{eq:SE}, the basis function is defined as:
\begin{equation}
    g(x) = \sigma\left(\frac{2}{\pi \xi^2}\right)^{1/4}\exp{\left(-\frac{x}{\xi}\right)^2}\;,
\end{equation}
and the convolution of two squared exponential kernels is
\begin{equation}
    \label{eqn:ksese}
    k_{1 \times 2}(x,x') = \sigma_1 \sigma_2 \left(\frac{2\, \xi_1 \xi_2 }{\xi_1^2+\xi_2^2}\right)^{\!1/2}\!\! \exp{\left(-\frac{\vert x-x' \vert^2}{\xi_1^2+\xi_2^2}\right)}\!.
\end{equation}

An important issue for multi-task GP is {`negative transfer'} and {`over-fitting'}. As discussed in \cite{Haridasu:2018gqm}, these occur when one data set is much more constraining than the others and so dominates the reconstruction. It is particularly problematic in regions where data are sparse. This is something which will be of importance with future data. For instance, higher quality and quantity data on $\fsig$ is likely to be released sooner than new data on $f$ or $\sig$. We leave the study of these caveats for future work.

\subsection{Mock data and analysis method}\label{sec:models}

The reference model we consider is flat $\Lambda$CDM, using the parameter values from the \emph{Planck} collaboration 2018 release \cite{Aghanim:2018eyx}. We set parameters to the mean values obtained via the robust combination of constraints from\footnote{See \emph{Planck} Legacy Archive: \url{https://pla.esac.esa.int/\#cosmology}.}: cosmic microwave background (CMB) base temperature, polarisation and lensing; baryon acoustic oscillations (BAO);  supernovae type Ia (SNIa). We then use \texttt{CLASS} \cite{2011JCAP...07..034B} to compute the redshift evolution of $f$, $\sigma_8$ and $f\sigma_8$ fiducials, in order to build our mock data sets. We generate simplified mocks as follows: 
\begin{itemize}
\item
Each mock contains 20 measurements uniformly distributed between redshift 0 and 2.
\item
We assume a constant relative error of 1\%, consistent with Stage IV expectations. 
\item
We define a multivariate Gaussian distribution with mean equal to the fiducial values of each growth function, and diagonal covariance  matrix defined by the fiducial values times the relative error. We do not incorporate cross-correlations in the mock data since these are too survey-specific. Mock data are then generated as random realisations of this distribution.
\end{itemize}
\vskip1mm
We optimise the hyperparameters, i.e. find their best fit, by minimising the log marginal likelihood in \autoref{eq:gp_log_marginal} applied to the data set. We use the stochastic minimiser `differential evolution' from the python scipy package\footnote{\url{https://docs.scipy.org/doc/scipy/reference/generated/scipy.optimize.differential_evolution.html}} in our GP code\footnote{ \url{https://gitlab.com/perenonlouis/GPreconstructions}}. The mean and  covariance of the final reconstruction are then obtained using this set of parameters in \autoref{eq:gpmean} and \autoref{eq:gpcov}. For the sampling of the posteriors of the hyperparameters via Markov chain Monte Carlo (MCMC) methods, we interface our GP code with \texttt{Cobaya\footnote{\url{https://cobaya.readthedocs.io/en/latest}} }\cite{Torrado:2020dgo}. This allows us to use a Metropolis Hasting sampler \cite{Lewis:2002ah,Lewis:2013hha} and the nested sampler \texttt{Polychord} \cite{Handley:2015fda,2015MNRAS.453.4384H}. We also interfaced the Affine-Invariant Ensembler Sampler \texttt{emcee\footnote{\url{https://emcee.readthedocs.io/en/stable/}} } \cite{ForemanMackey:2012ig}. All our statistical analyses and post processing of our results are done with \texttt{GetDist\footnote{\url{https://getdist.readthedocs.io/en/latest/}}} \cite{Lewis:2019xzd}.

\section{Performance and robustness}\label{sec:perfo}

Here we investigate the performance of multi-task GP compared to single-task and make some tests of the results.

\subsection{`Sharing is caring'}\label{sec:sharing}

\begin{figure}[!]
\begin{center}
\includegraphics[trim=0 77 0 0,clip,width=0.92\linewidth]{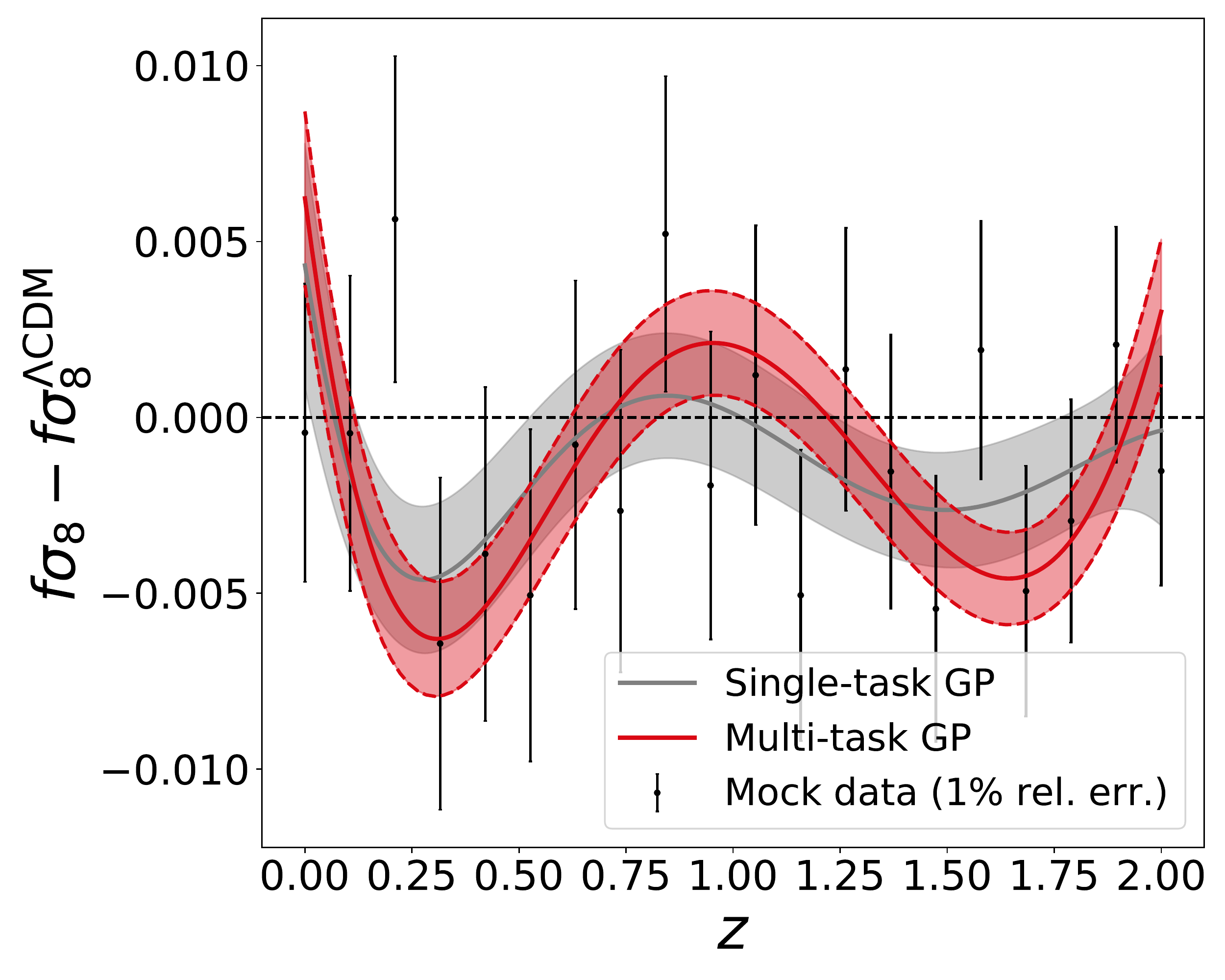}\\
\hskip3mm\includegraphics[trim=0 77 0 0,clip,width=0.88\linewidth]{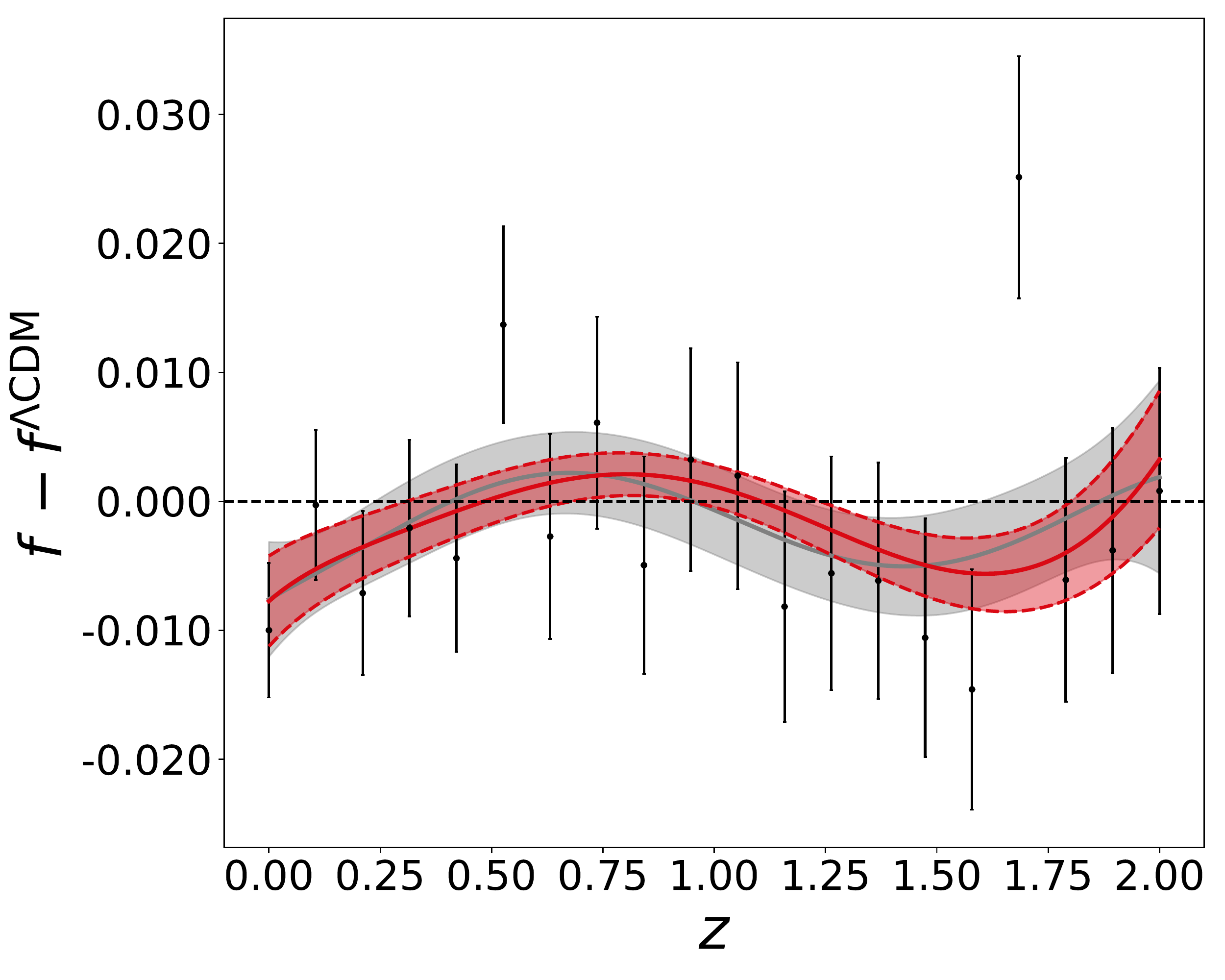}
\includegraphics[width=0.92\linewidth]{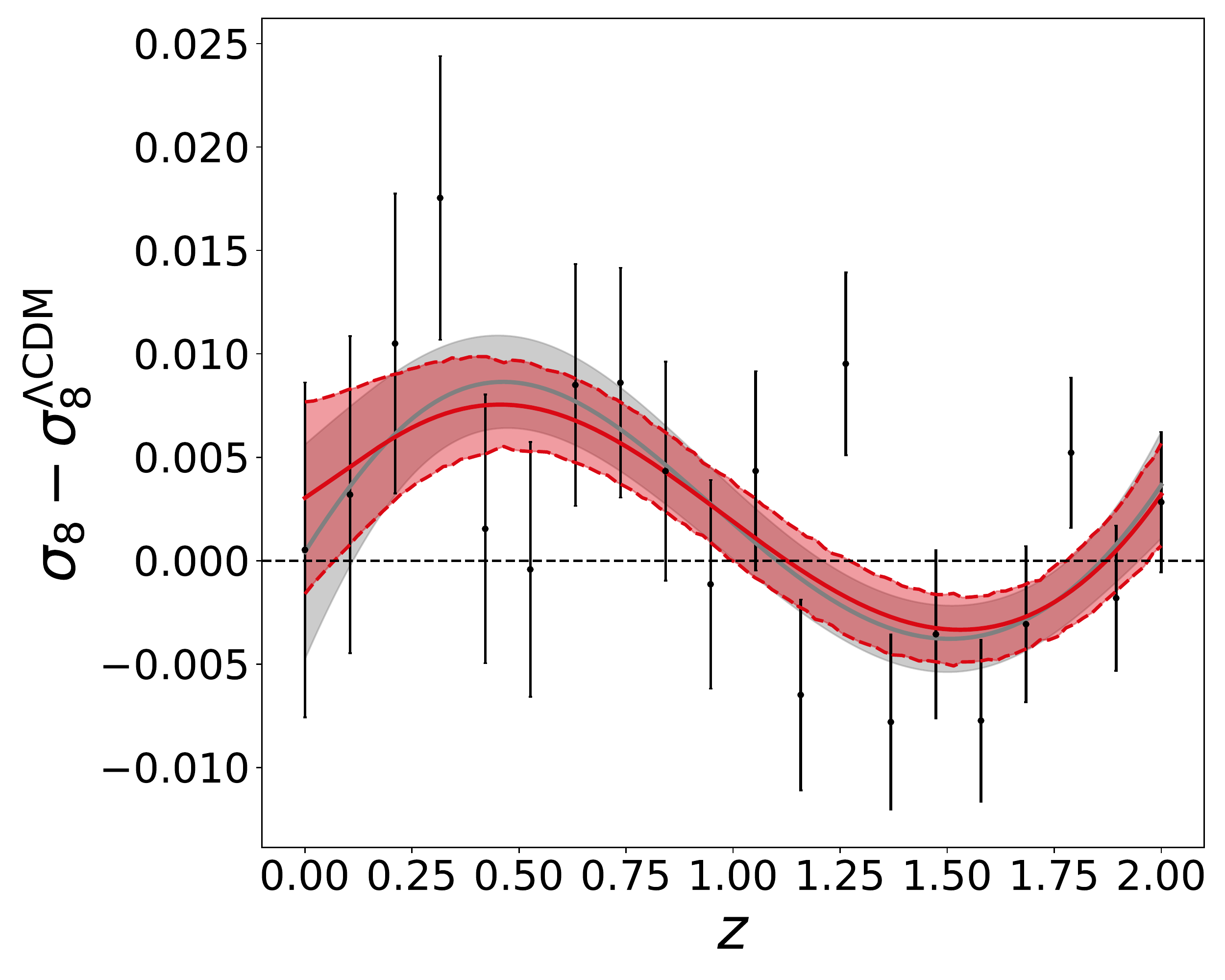}
\caption{Optimised single- (grey) and multi-task (red) reconstructions of $\fsig$, $f$ and $\sig$ with 1\% relative error mocks, using $\Lambda$CDM fiducial constrained by CMB T+P+L, BAO, SNIa (see \autoref{sec:models}), for squared-exponential kernel and null mean prior. Shading indicates 68\% confidence intervals about the mean (solid curve). The multi-task GP reconstructs the three growth functions simultaneously by minimising a common likelihood (see \autoref{sec:mtgp}) -- thus `sharing' information between the three reconstructions. As a result the single- and multi-task predictions do not match exactly: multi-task reconstruction for a given function benefits from the constraints on the other functions, thereby decreasing the error on all the reconstructions. Quantitatively, the multi-task GP provides the best fit to the mock data.}
\label{fig:stgp_vs_mtgp_plotz}
\end{center}
\end{figure}

We use a squared-exponential kernel \autoref{eq:SE} and a null mean prior as the base assumptions for the single- and multi-task GP. Note that the reconstructions are robust under the change of these assumptions for reasonable choices. This is studied in \autoref{sec:assumptions}. 

An overlay of the single- and multi-task reconstructions is shown in \autoref{fig:stgp_vs_mtgp_plotz}, for the generation of a single mock. At this level of precision, the reconstructions are very sensitive to the distribution of the data around the fiducial. We discuss  the deviations of the reconstructions in \autoref{sec:deviations}. We concentrate for now on the differences between single- and multi-task reconstructions.

\autoref{fig:stgp_vs_mtgp_plotz} further shows that multi-task GP increases the precision of reconstructions compared to single-task GP. It also changes the redshift trends relative to the single-task case. Although the error for the multi-task GP is in general reduced, this is not guaranteed at all redshifts. The changes in the redshift trends and error sizes are a clear manifestation of the sharing of information between the reconstructions of functions in the multi-task case. Quantitatively, we find that the multi-task log marginal likelihood is larger than the sum of the single-task ones.

\begin{figure*}[!]
\begin{center}
\includegraphics[scale=0.58]{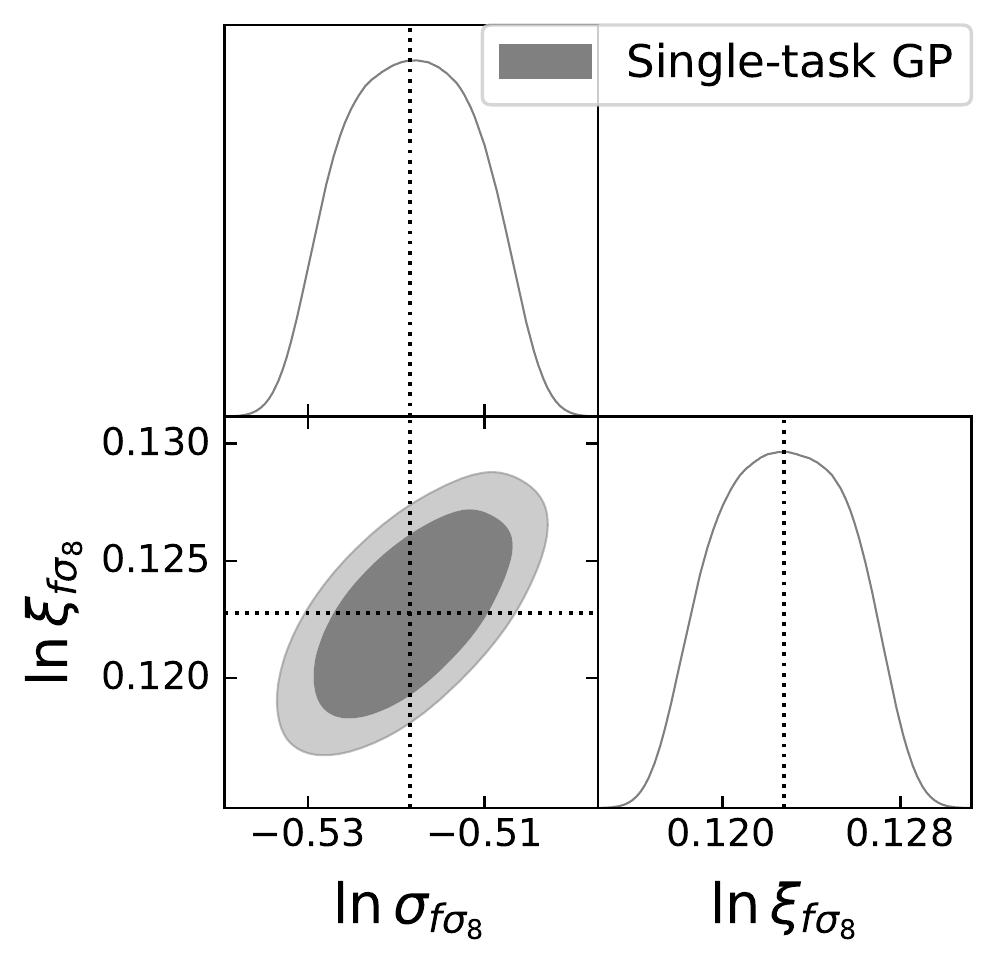}
\includegraphics[scale=0.58]{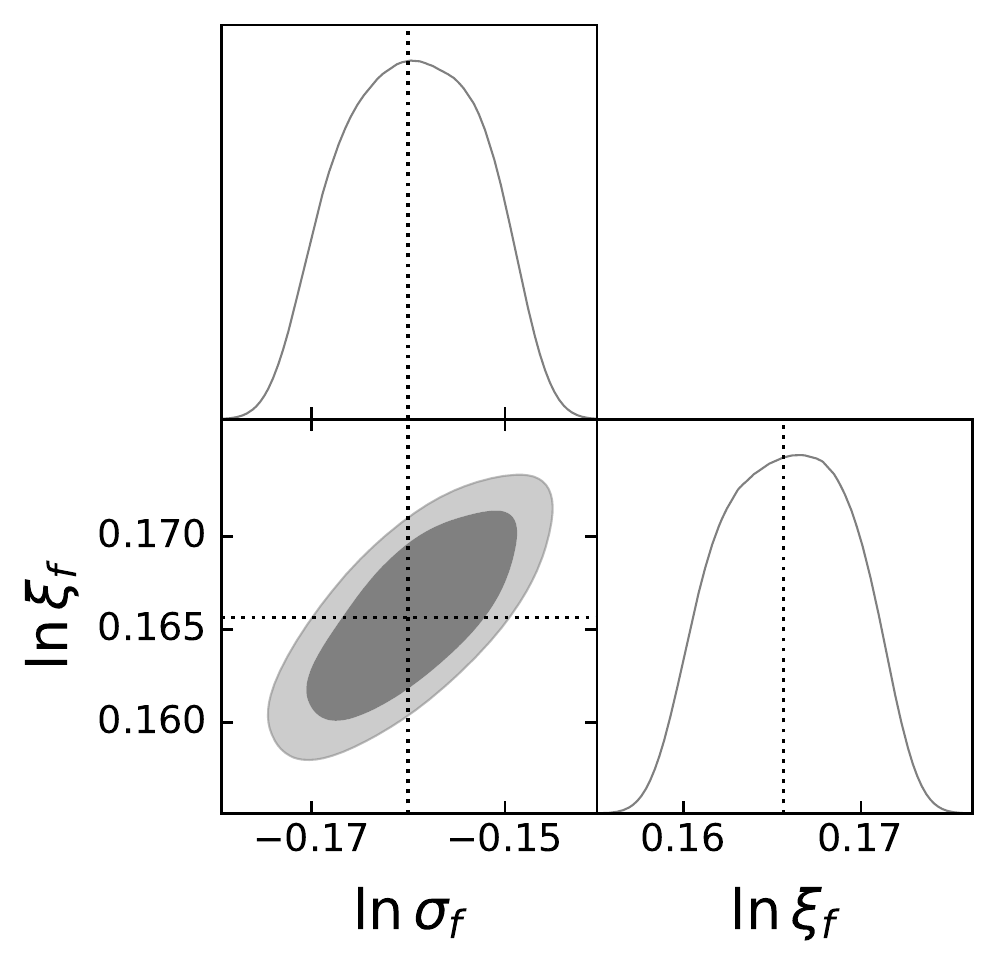}
\includegraphics[scale=0.58]{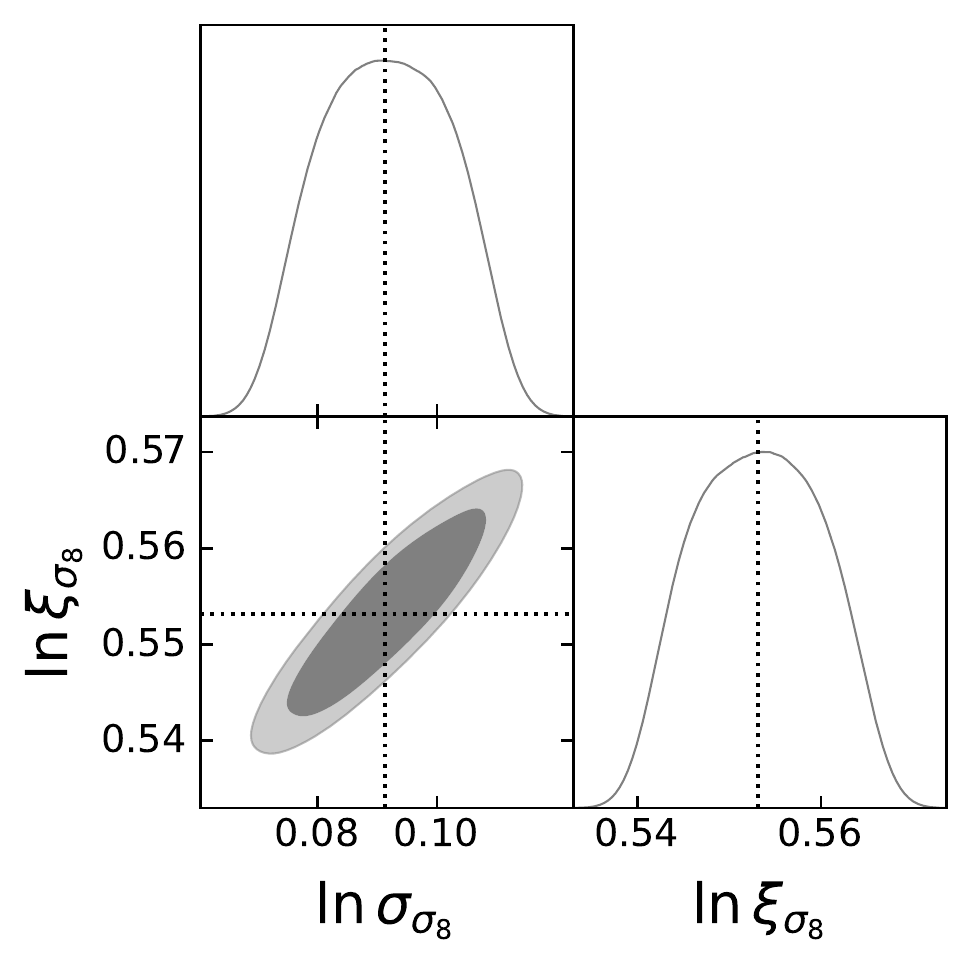}\vskip4mm
\includegraphics[scale=0.45]{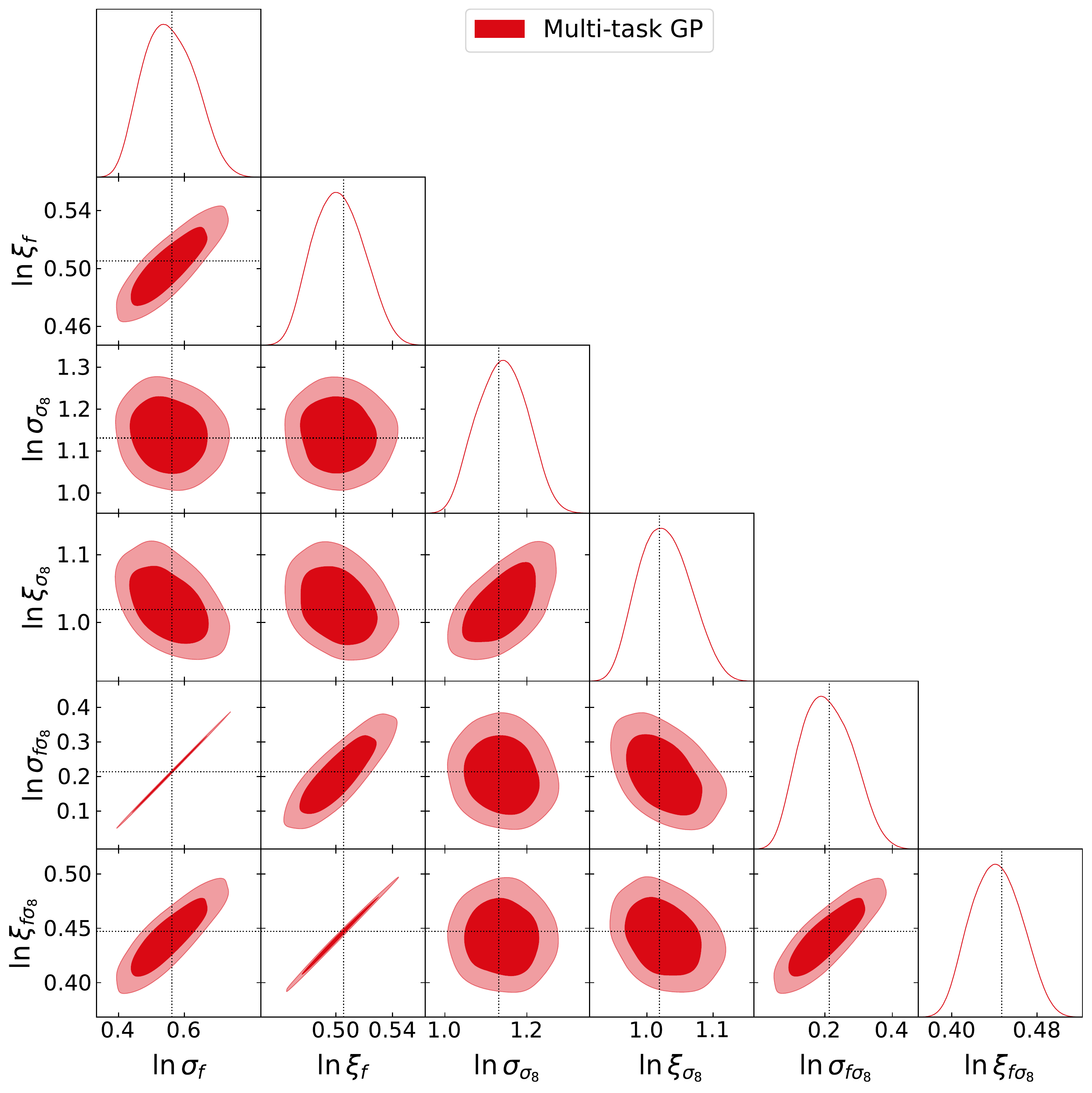}
\caption{Posterior distributions from the MCMC sampling of the hyperparameters for the single-task (grey contours) and multi-task (red contours) reconstructions shown in \autoref{fig:stgp_vs_mtgp_plotz}. Dashed lines indicate optimised values. The transfer of information between reconstructions in the multi-task case adds correlations between hyperparameters of each kernel and the hyperparameter posteriors become much wider. It also increases values of all hyperparameters relative to single-task, so that the reconstructions will be more `wiggly', as seen in \autoref{fig:stgp_vs_mtgp_plotz}.}
\label{fig:stgp_vs_mtgp_contours}
\end{center}
\end{figure*}

A complementary approach to optimising the GP is to reconstruct the full posterior distributions of the hyperparameters. This allows in particular the possibility to study the degeneracies between them. Sampling the space of hyperparameters from the log marginal likelihood in \autoref{eq:gp_log_marginal} can be achieved using Monte Carlo Markov Chains (MCMC). The multi-task case is more difficult to sample because of non-global minima. We found that a nested sampler is more robust than a Metropolis-Hasting sampler. For the same reason, optimising the hyperparameters with a stochastic minimiser is safer than any method based on gradient descents or similar. The results of the MCMC sampling corresponding to the GP of \autoref{fig:stgp_vs_mtgp_plotz} are shown in \autoref{fig:stgp_vs_mtgp_contours}. Sampling the hyperparameters also allows us to derive the marginalised reconstruction. We verify that the Gaussianity of the mocks renders each best-fit value of the reconstructions to be very close to the maximum of their marginalised posterior. This implies that the optimised reconstructions are virtually the same as the marginalised ones \cite{Seikel:2012uu}.

\autoref{fig:stgp_vs_mtgp_contours} shows that the 2-dimensional posteriors between two hyperparameters of the same kernel display correlations whatever the approach. The transfer of information between the reconstructions in the multi-task case results in correlations between hyperparameters of each kernel. The hyperparameter posteriors are therefore much wider. This means that previously ruled out values of hyperparameters for one of the functions are now allowed, as long as the others change accordingly. This transfer of information also has the effect of increasing the values of all hyperparameters relative to the single-task case. Interestingly, this implies that one effect of the convolutions in the multi-task kernel is to increase the correlation of the data as seen by the GP. The hyperparameters for each reconstruction are larger and therefore they will be more `wiggly'. This is consistent with what seen in \autoref{fig:stgp_vs_mtgp_plotz}. 

\begin{figure}[!]
\begin{center}
\includegraphics[width=0.92\linewidth]{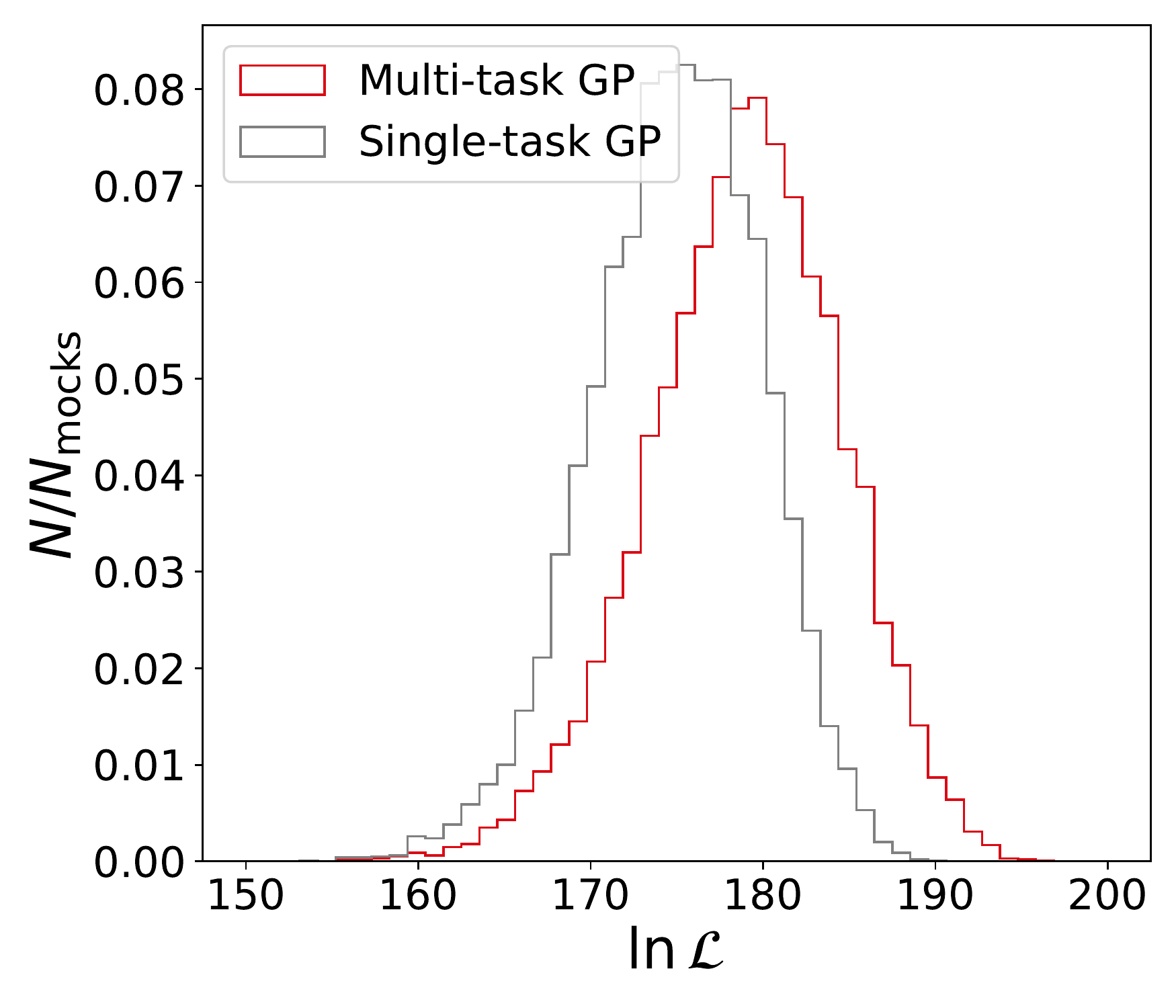}
\caption{Distribution of best-fit values for the multi-task log marginal likelihood (red) and the sum of single-task log marginal likelihoods (grey) from MC generation of {$10^4$} mocks. The sum of the single-task log marginal likelihoods is smaller than the multi-task log marginal likelihood, indicating that the multi-task approach fits the data better.}
\label{fig:stgp_vs_mtgp_chi2s_x}
\end{center}
\end{figure}

A Monte Carlo generation of $10^4$ mocks confirms that the above conclusions are robust. We find that the distribution of best-fit multi-task log marginal likelihood is higher than the sum of the three single-task ones, as shown in \autoref{fig:stgp_vs_mtgp_chi2s_x}. The multi-task thus performs a better fit to the union of the three growth data sets. This highlights an important point: the multi-task approach reduces the width of the confidence interval around the mean, but it also changes its redshift evolution. The single-task approach reconstructs more faithfully the redshift dependence of each data set -- simply because it is oblivious to the data on the other functions. Since the log marginal likelihood remains better for the multi-task case, this means that the reduction of the confidence interval width of the reconstruction dominates over the change of redshift evolution in the goodness-of-fit.

\subsection{Assumptions}\label{sec:assumptions}

The first step to evaluate the robustness of GP reconstructions  is to compare predictions with different kernels. The more dependent the results are on this choice, the less  model-independent is the reconstruction. For the growth of structure, we reconstruct underlying smooth functions. We therefore consider only stationary kernels. For these tests, we also concentrate on the single-task reconstruction of $\fsig$. 

\begin{figure}[!h]
\begin{center}
\includegraphics[trim=0 77 0 0,clip,width=0.9\linewidth]{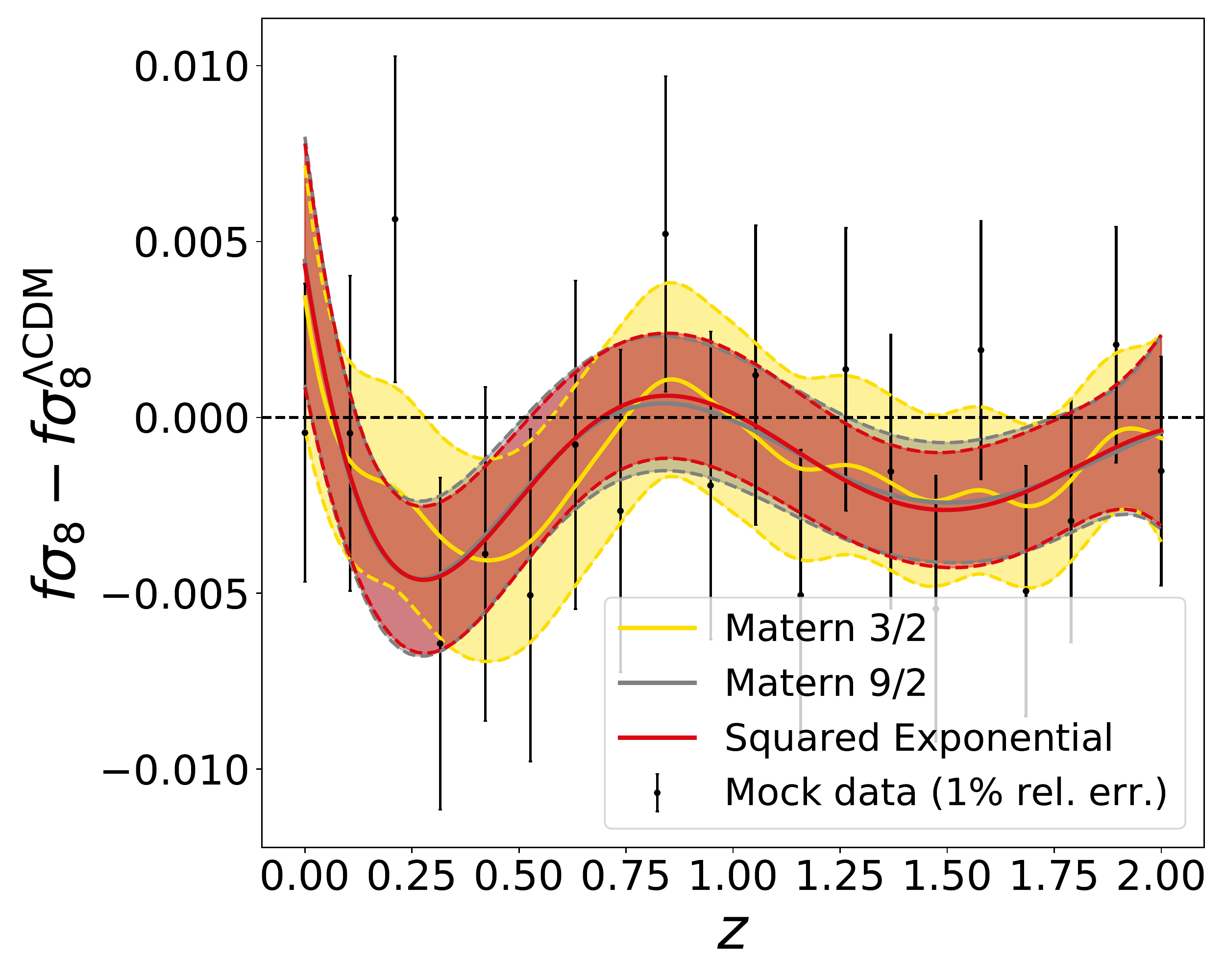}\\
\includegraphics[trim=0 0 0 0,clip,width=0.9\linewidth]{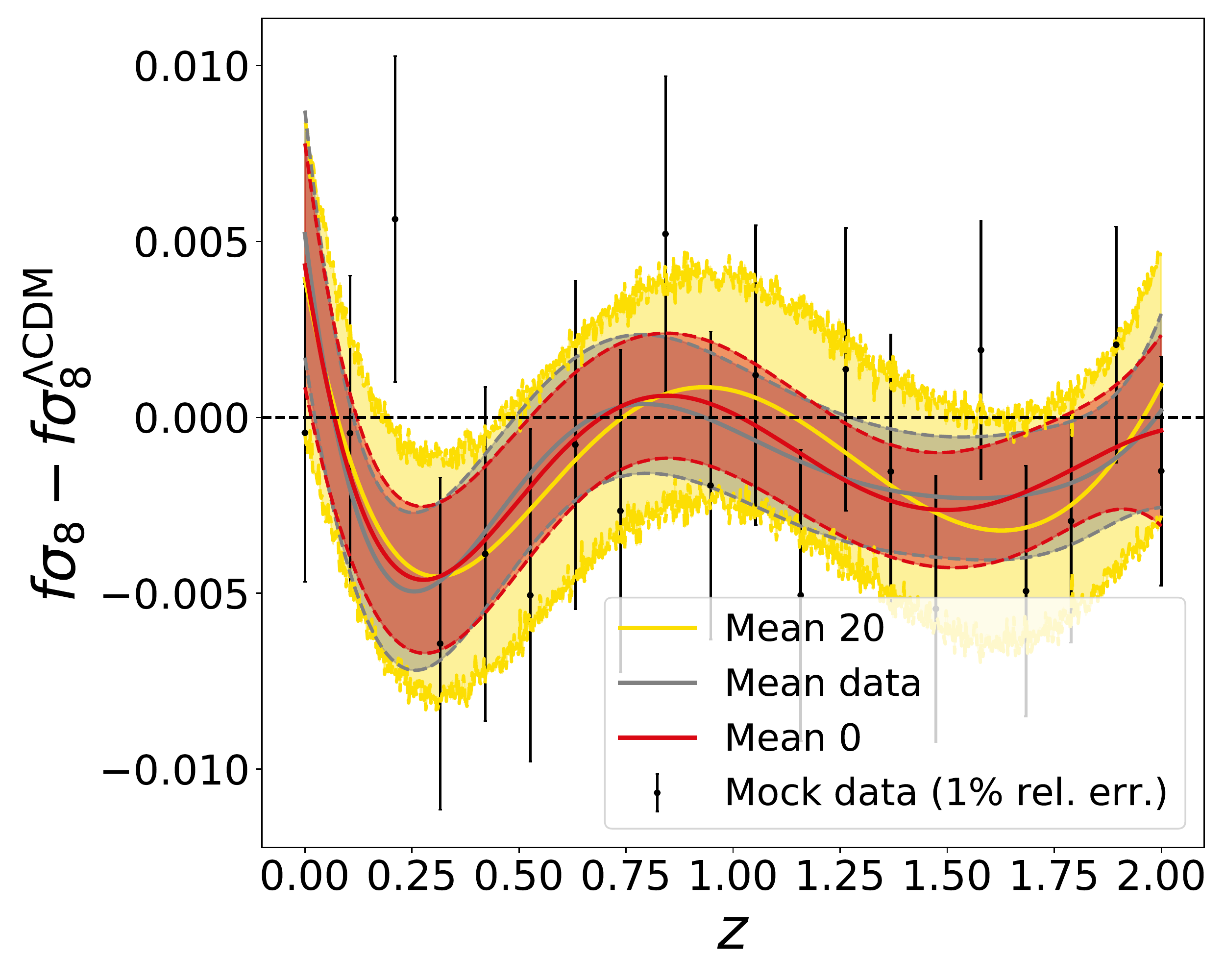}
\caption{Optimised single-task reconstructions of $\fsig$ with 1\% relative error mock, $\Lambda$CDM fiducial constrained by CMB T+P+L, BAO, SNIa. Shading indicates 68\% confidence intervals about the mean (solid curves). Reconstructions are shown with different kernels (top) and different mean priors (bottom). The Matern kernel with $\nu =3/2$ produces jagged reconstructions, while squared exponential and Matern $\nu =9/2$ give very similar results. 
The unreasonable choice of mean prior shows large and noisy errors, while sensible mean priors  produce similar reconstructions.}
\label{fig:assumptions}
\end{center}
\end{figure}

We test all the common stationary and non-periodic kernels: exponential, squared exponential, gamma exponential, the Matern class and rational quadratic (see \cite{Rasmussen}). We find that the exponential, gamma exponential, and Matern with $\nu < 5/2$ produce jagged reconstructions, as shown in \autoref{fig:assumptions}. These are discarded  on the basis of the smoothness requirement. Reconstructions performed with the other kernels all give very similar results. The rational quadratic kernel produces virtually the same reconstruction and goodness of fit as the squared exponential. This kernel can be understood as a scale mixture, i.e. an infinite sum, of squared-exponential kernels with different correlation lengths \cite{Rasmussen}. It also has an extra  hyperparameter. The absence of difference between  reconstructions with these kernels thus highlights that adding hyperparameter freedom does not produce new features. 

The Matern class  reconstructions produce errors that increase slightly as $\nu$ decreases. In the limit $\nu \rightarrow \infty$ the Matern kernel recovers the squared exponential. Not surprisingly, the case $\nu=9/2$ is very similar to the squared exponential, as illustrated in \autoref{fig:assumptions}. On the other hand, the smaller is $\nu$ the more sensitive will this kernel be to the noise of the data. This is why a reconstruction with $\nu=3/2$ shows short wiggles on top of the evolution reconstructed by the squared exponential (see \autoref{fig:assumptions}). 
 
In conclusion, from a GP point of view the simplest kernel, the squared exponential, captures all the information about the smooth evolution underlying the data. We  do not find any need  to consider combinations of kernels.  

The next step is to look at the other input defining a GP, i.e. its mean prior. The influence of the mean prior becomes important when the latter is either far away from the data or when the data displays gaps in redshift. The former case should not occur here, given reasonable user choices. The latter could be relevant in the future, depending on survey choices. To see how sensitive the input is with our mock set-up, we compare two common choices of mean prior, i.e. equal to zero or to the mean of the data, and an unreasonable choice, equal to 20. The results are shown in \autoref{fig:assumptions}. 

The unreasonable choice displays large and noisy errors. Nonetheless, the precision of the data is such that the reconstruction does not go to the mean prior in between any neighbouring measurements,  even though the mean prior is chosen significantly far away from the data. For mean prior equal to the mean of the data, the maximum log marginal likelihood is slightly enhanced relative to the null mean prior. However, the reconstruction $\chi^2$ is slightly increased. For real future data, it could be more meaningful to use the mean of the data as mean prior. 

Alternatively, one could input the standard model prediction as mean prior. Here, this is exactly the fiducial, and it produces a flat reconstruction around the fiducial, i.e. with correlation length hyperparameter $\xi$ diverging to infinity. Overall, we avoid a mean prior that is dependent on each mock, to ensure that each mock is treated on the same ground during the MC generations. 
We conclude that the null mean prior is a sound benchmark.

Regarding the priors on the range of the hyperparameters, we choose them large enough that they have no effect on the reconstructions. We thus ensure that the optimised value of a hyperparameter is not  its maximum or minimum allowed. Note that these considerations do not apply when using currently available growth data (see \autoref{sec:current}), highlighting how the precision of observations is a crucial ingredient to apply machine learning tools to cosmological analyses. 

We also verified that marginalising over the hyperparameters produces virtually the same reconstructions as when optimised. This is expected, since the precision of the mock data and its Gaussian distribution around the fiducial produces posteriors that are very close to Gaussian, as shown in \autoref{fig:stgp_vs_mtgp_contours}. 

In conclusion, this analysis shows that at this level of precision of the mock data, the reconstructions are fully driven by data, indicating that our comparisons of the single- and multi-task approaches are robust. 

\subsection{Bias and deviations}\label{sec:deviations}

\begin{figure}[!]
\begin{center}
\includegraphics[trim=0 0 0 00,clip,width=0.9\linewidth]{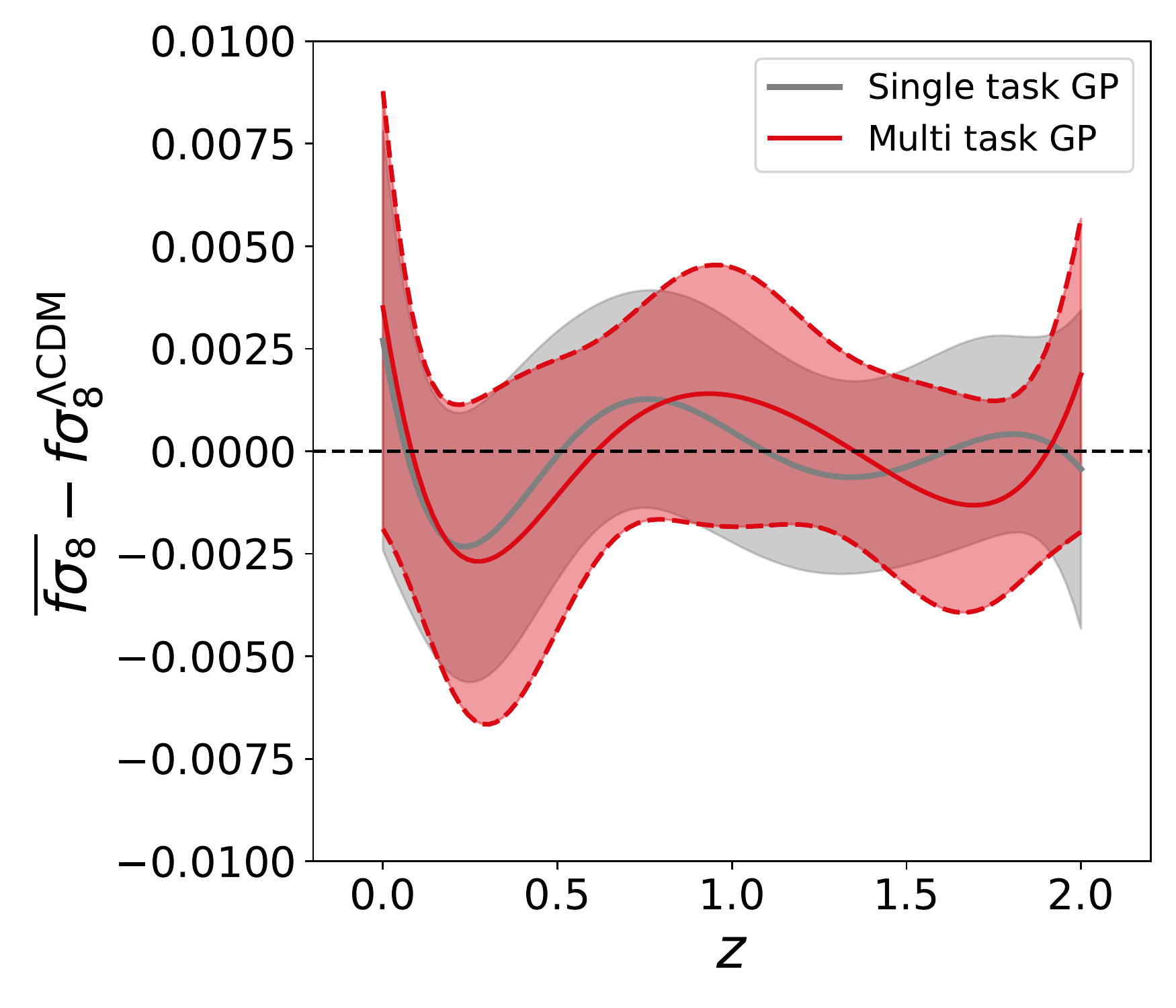}
\caption{Average distribution over $10^4$ mocks of the $\fsig$ reconstructions, computed by drawing 1000 realisations of each optimised GP, from the multivariate Gaussian built with their mean and covariance. We concatenate all 1000$\times${$10^4$} draws. The average fluctuates around the fiducial function, which is well recovered within the error.}
\label{fig:bias}
\end{center}
\end{figure}

Reconstructions can deviate quite significantly from their fiducials given the precision and distribution of the mock data. How likely are these departures and are any biases hidden? We run a series of tests to answer these questions.

First, we check for any overall bias in our reconstruction by performing the equivalent of an `ensemble average' of the reconstructed functions, over the $10^4$ mocks. To estimate the average distribution analytically would require the computation of a mixture of multivariate Gaussian distributions (i.e.\ the outputs from each GP optimisation) which has no closed form. Thus, we compute this average distribution numerically by drawing 1000 realisations of each optimised GP -- from the multivariate Gaussian built with their mean and covariance -- and we concatenate all 1000$\times${$10^4$} draws. We then derive the overall mean and the 68\% limits of the resulting distribution. The results for $\fsig$ in are displayed \autoref{fig:bias}. We note that the average fluctuates around the fiducial function -- but nevertheless the fiducial is well recovered within the error and no significant biases can be seen. We also verified that the distributions at a given redshift are Gaussian. The same conclusions hold for the other two functions.

\begin{figure}[!]
\begin{center}
\includegraphics[width=0.68\linewidth]{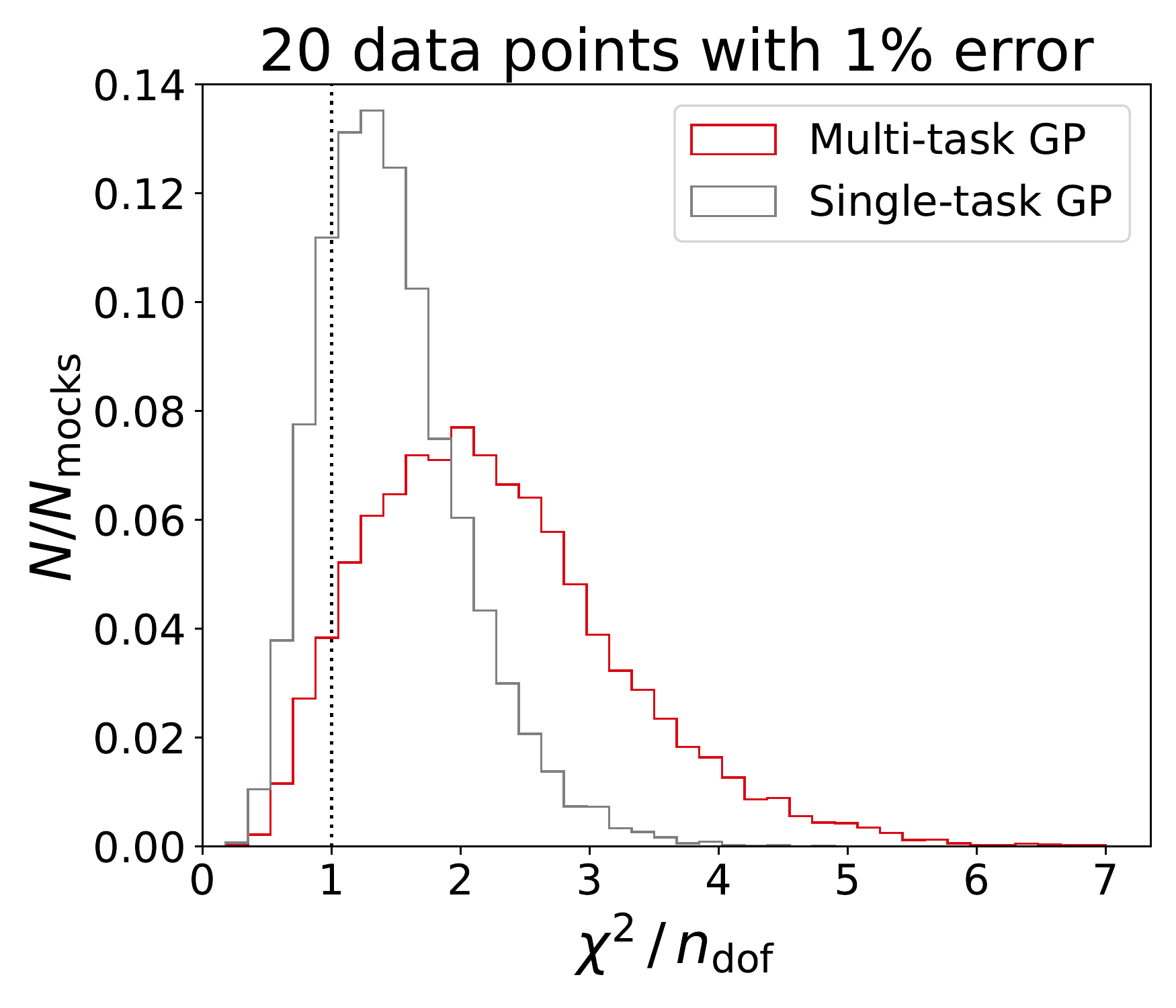}
\includegraphics[width=0.68\linewidth]{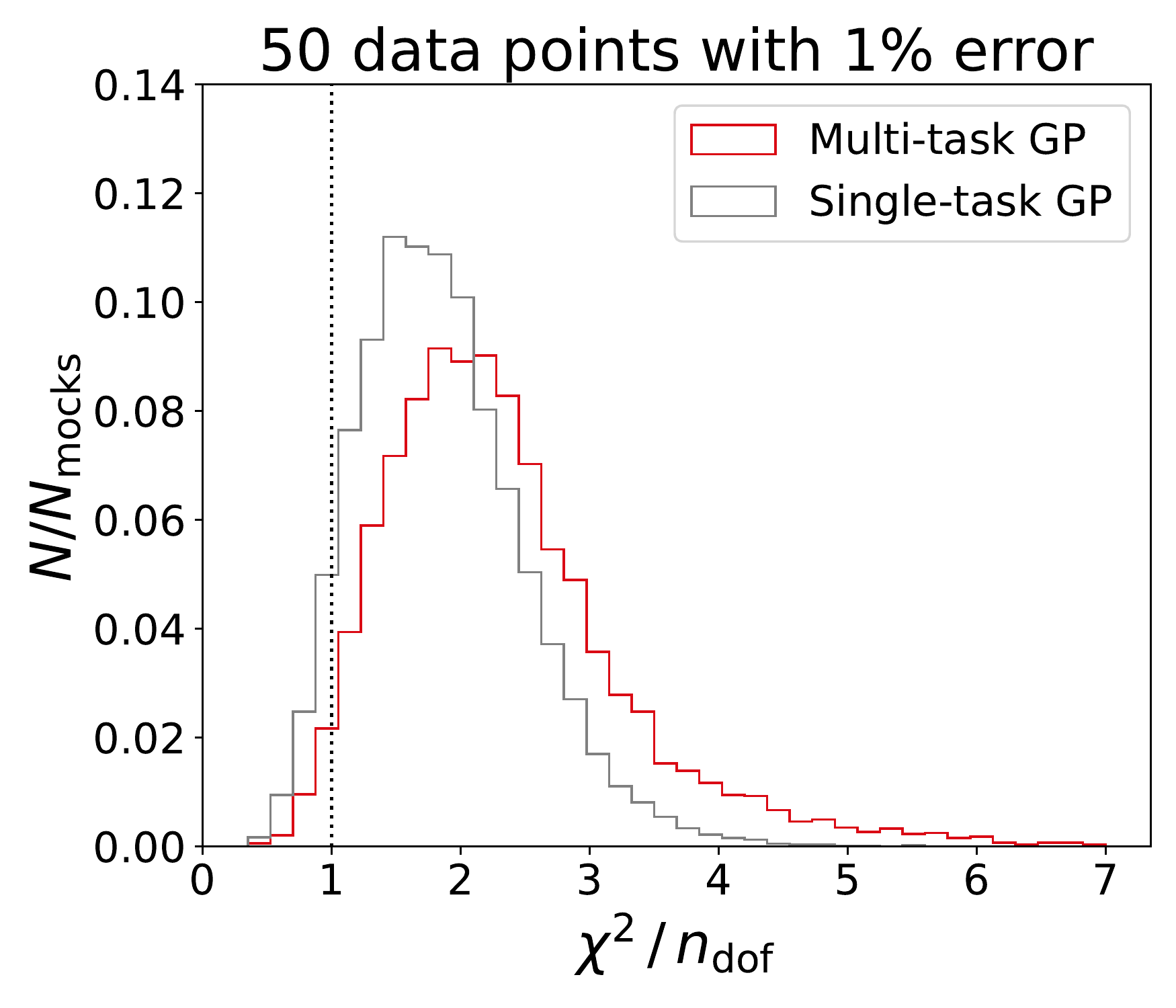}
\includegraphics[width=0.68\linewidth]{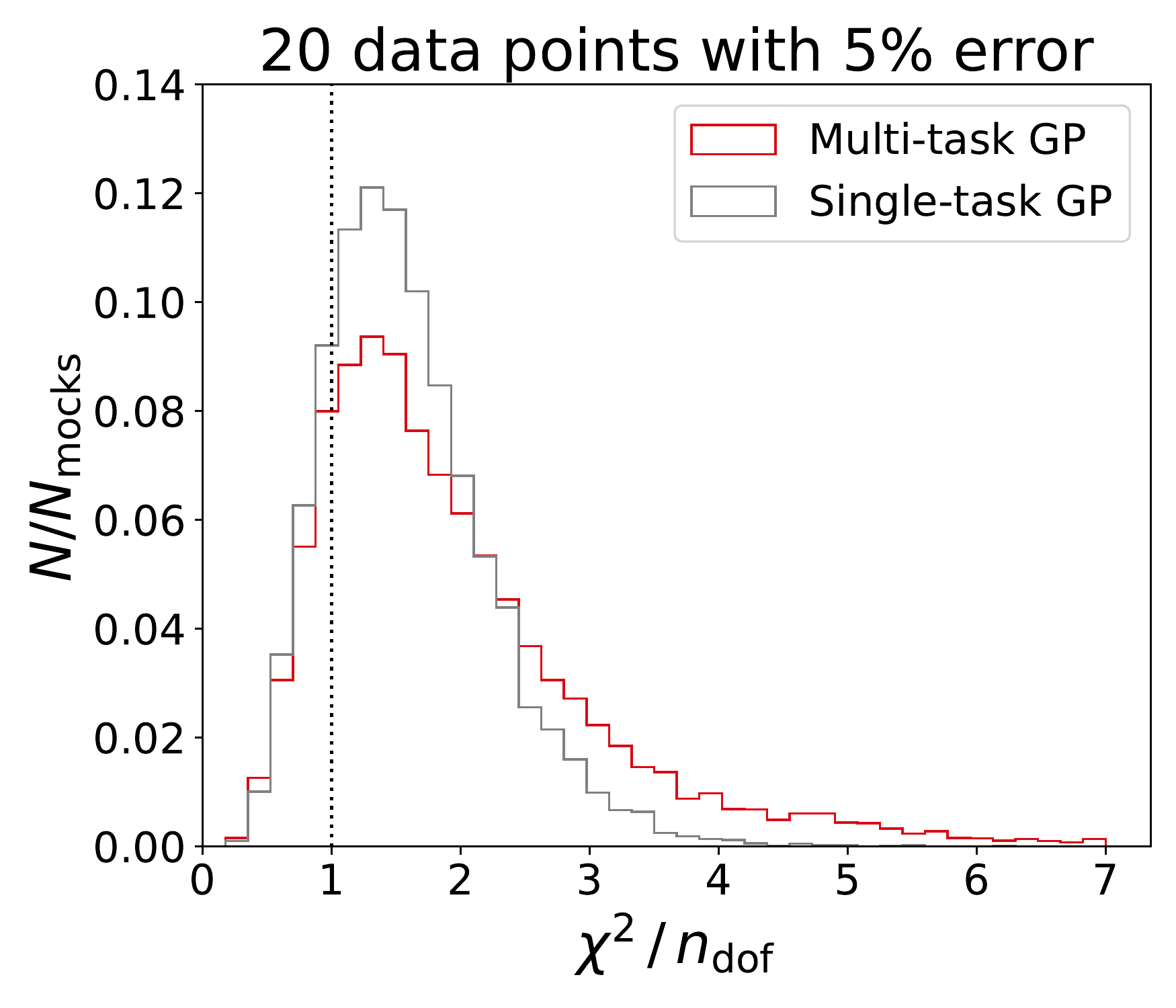}
\caption{Distributions of the reduced $\chi^2/n_{\rm dof}$ (see \autoref{eq:chi2}) of the combined $\fsig\;, f\;, \sig$ reconstructions for the $10^4$ mocks. We vary some characteristics of the mocks: 20 points uniformly distributed between redshift 0 and 2 with 1\% relative error (top); increased number of data points to 50 (middle); decreased precision of the data to 5\% for 20 points (bottom). The distributions of the reduced $\chi^2/n_{\rm dof}$ change depending on the configurations. This highlights that the sensitivity of GP's to the probability distribution of the data depends on the method -- single-task or multi-task -- and on the characteristics of the mock data itself.}
\label{fig:stgp_vs_mtgp_chi2s}
\end{center}
\end{figure}

The second check is to quantify how significantly  the reconstructions deviate from the fiducial. In order to do so, we exploit the fact that we know the true underlying model in our case study, i.e. the fiducial. We define a hybrid $\chi^2$ between the optimised reconstructions and the fiducial as 
\begin{equation}\label{eq:chi2}
    \chi^2 = \sum_{g}\sum_{ij} \left(Y^i_{\rm fid}-Y^i_{g}\right) C^{-1}_{ij,{g}}\left(Y^j_{\rm fid}-Y^j_{g}\right)\, ,
\end{equation}
summed over the three growth reconstructions $g=\lbrace{f,\, \sigma_8,\, f\sigma_8 \rbrace}$. For each reconstruction, the $\chi^2$ is computed for a vector $\bm{Y}$ of 100 redshifts uniformly distributed between 0 and 2. $\bm{C}_{g}^{-1}$ is the corresponding inverse diagonal covariance matrix of the reconstruction. The number of degrees of freedom is therefore $n_{\rm dof}=3 \times 100$. The results are displayed in \autoref{fig:stgp_vs_mtgp_chi2s}.

We find that the reduced $\chi^2$ of the GP reconstructions is not distributed around unity (see top plot). This shows that given the characteristics of the mocks -- relative error, Gaussian probability distribution, number of data -- the reconstructions are too sensitive to the probability distribution of the data and do not recover the fiducial efficiently. In addition, the multi-task distribution is shifted to higher values and is flattened relative to the single-task. This is the result of the tighter confidence intervals produced by the former and not to any particular bias, since we found the reconstructions to be Gaussian distributed on average around the fiducial.

The third check is to test how modifying the characteristics of the mocks influences the previous conclusions. On the one hand, adding data points in the same redshift range (\autoref{fig:stgp_vs_mtgp_chi2s}, middle) does not affect significantly the distribution of reduced $\chi^2$ for the multi-task case. Interestingly, the reduced $\chi^2$ of the single-case is pulled towards reproducing that of multi-task. On the other hand, lowering the relative error of the data shifts the distributions toward a better fiducial recovery.

\begin{figure}[!]
\begin{center}
\includegraphics[width=0.7\linewidth]{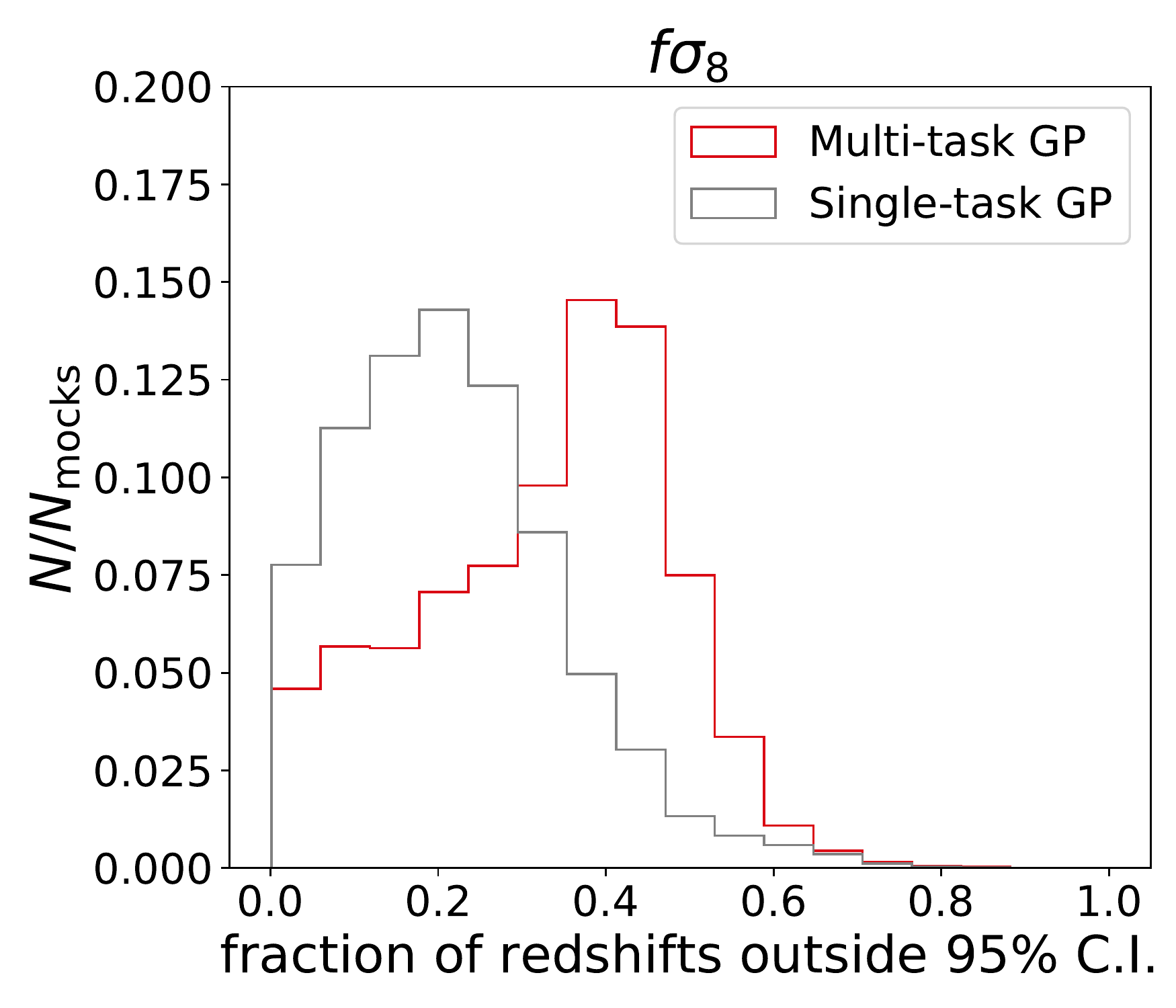}
\includegraphics[width=0.7\linewidth]{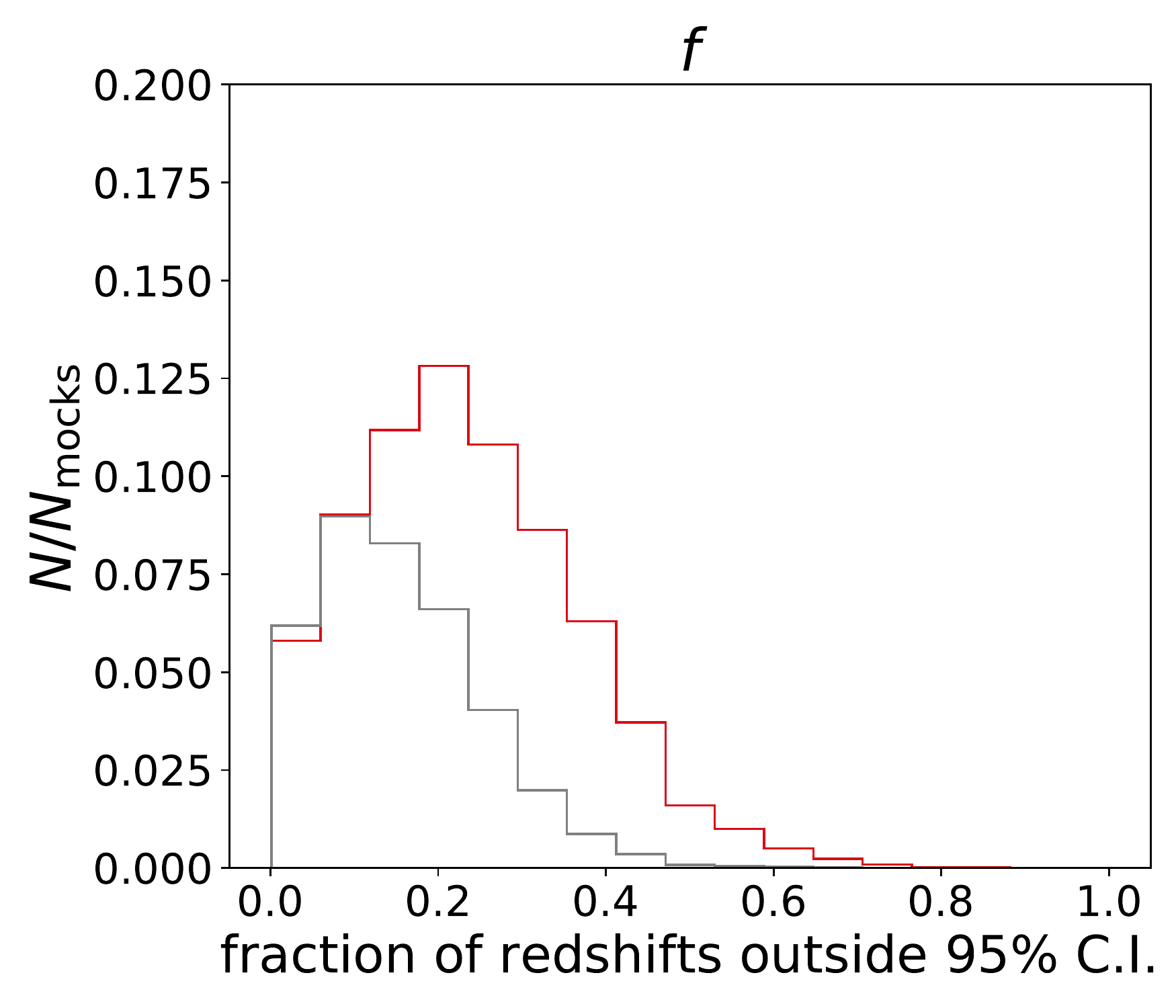}
\includegraphics[width=0.7\linewidth]{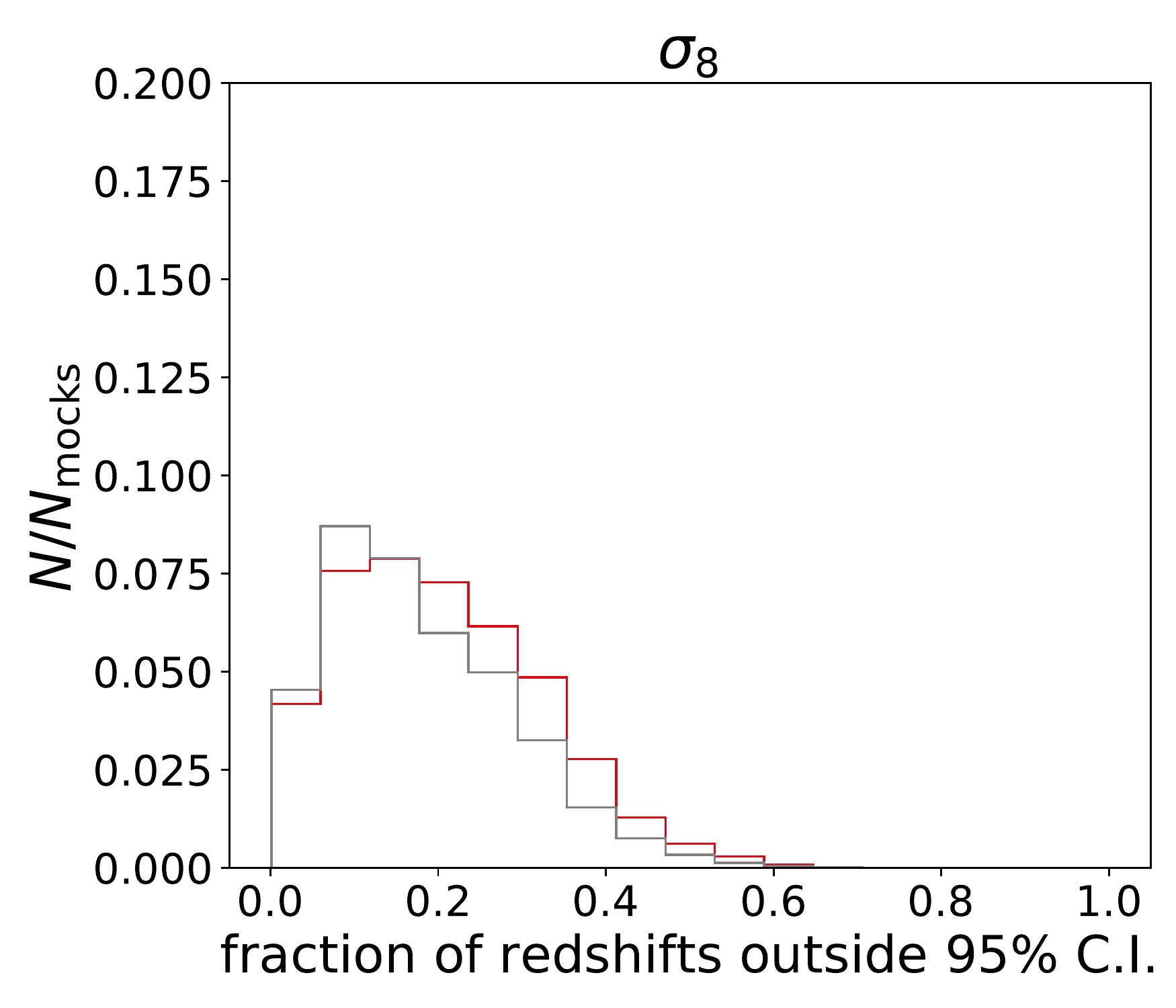}
\caption{Distribution of the fraction of the redshift range over which the fiducial falls outside of the 95\% confidence interval of the reconstructions. The distribution for each  reconstruction is computed over the $10^4$ mocks for 100 points uniformly distributed between redshift 0 and 2. The distributions depend significantly the function considered.}
\label{fig:stgp_vs_mtgp_counts}
\end{center}
\end{figure}

The fourth test we conduct is to analyse quantitatively how much  the reconstructions deviate from the fiducial. To this end, we compute for each mock the fraction of the redshift range over which the fiducial falls outside of the 95\% confidence interval of the reconstructions \cite{Seikel:2012uu}. We display the distribution of this fraction over the $10^4$ mocks for our three reconstructed functions in \autoref{fig:stgp_vs_mtgp_counts}. The results show a trend similar to our previous $\chi^2$ test -- i.e., the tighter multi-task errors exclude the fiducial function more often. Another important result highlighted by these plots is that the distributions depend significantly on the function considered, and hence on the characteristics of the mocks. For instance, the redshift evolution of the function has an importance since we consider relative errors to construct the mocks.

The tests in this section highlight some of the subtleties associated with GP regressions. Even when the reconstruction is not biased on average, the fact that it performs `too well' in some cases (i.e. has stringent error bars) can complicate the assessment of whether it excludes a given cosmological model or not. Moreover, as we showed above, such effects can depend on the specific choice of mocks, hence ultimately on the intrinsic properties of the data (noise amplitude, redshift dependence, etc.) to which we apply the GP reconstruction. A full diagnostic of such effects will be the topic of an in-depth future work.

\section{Applications to current data}\label{sec:current}

\begin{figure}[!]
\begin{center}
\includegraphics[width=0.9\linewidth]{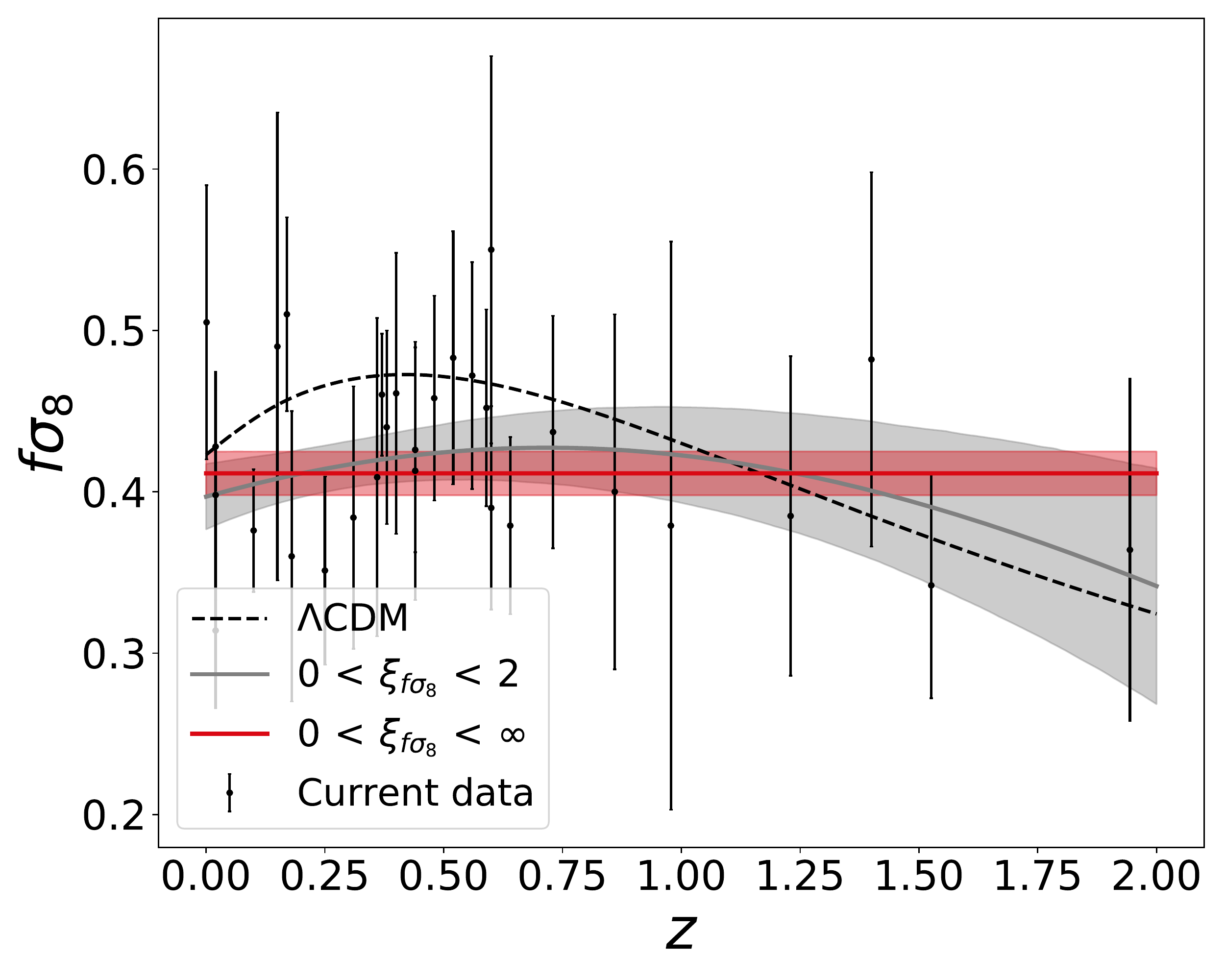}
\includegraphics[width=0.9\linewidth]{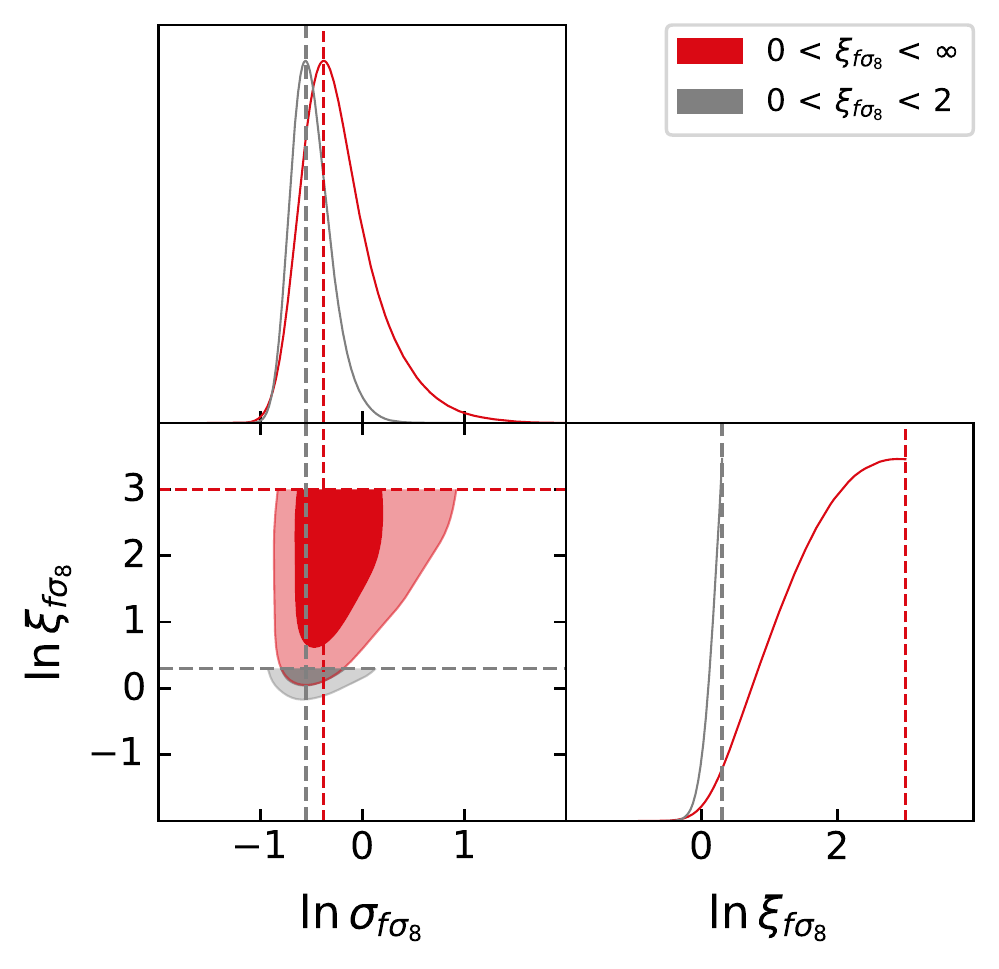}
\caption{{\it Top:} Optimised single-task reconstructions of current $\fsig$ data with squared-exponential kernel and null mean prior. Upper prior bounds on correlation length hyperparameter $\xi_{\fsig}$ are either restricted the maximum redshift range of the data (grey) or unbounded (red). Shading indicates 68\% confidence intervals. Dashed curve is the $\Lambda$CDM fiducial. {\it Bottom:} The corresponding MCMC sampling of posterior distributions on the hyperparameters. Shading shows 68\% and 95\% confidence intervals. Dashed lines indicate optimised values of hyperparameters. If $\xi_{\fsig}$ is unbounded from above then its best fit wants to be infinite which produces a flat reconstruction; in practice it will be very close to its maximum range prior.}
\label{fig:reco_current}
\end{center}
\end{figure}

To conclude our analysis we investigate reconstructions of current growth data. We consider first the compilation of measurements of $\fsig$ in \cite{Perenon:2019dpc} (updated using \cite{Nesseris:2017vor}). By using this data from different and non-overlapping galaxy surveys, we avoid possible double counting due to overlapping areas or different releases of the same survey. (This differs from the approach of \cite{Yin:2018mvu,Benisty:2020kdt,Li:2019nux}, where an extended set of RSD measurements is considered.) 

The result (with squared-exponential kernel) is shown in \autoref{fig:reco_current}. It is apparent that current RSD data does not allow for a robust GP reconstruction, which depends on the allowed range for the correlation hyperparameter $\xi_{\fsig}$ (red and grey shadings). This dependence persists for different choices of mean prior, including the standard model, and for different kernels.

Allowing the range of $\xi_{\fsig}$ to be very large leads to a posterior that is unbounded from above when explored via MCMC, as shown in the bottom panel. The best-fit value for $\xi_{\fsig}$ tends to infinity and in practice will always be close to the maximum range allowed. This is why the reconstruction is flat (red shading). The error on the reconstruction is then carried solely by the signal variance $\sigma_{\fsig}$, whose 1-dimensional posterior distribution is on the contrary well defined.

This illustrates that current RSD data is simply too noisy for a GP reconstruction independent of assumptions. The data is effectively seen as random noise by the GP.

If we use the extended data set of \cite{Yin:2018mvu,Benisty:2020kdt,Li:2019nux}, the same problem arises. The only way we can reproduce a growth reconstruction similar to \cite{Zhang:2018gjb,Yin:2018mvu,Li:2019nux,Benisty:2020kdt} is to require the range of $\xi_{\fsig}$ to be smaller than the redshift range of the data. This forces the best-fit value of $\xi_{\fsig}$ to be that of the chosen maximum range, as shown in the grey shadings in \autoref{fig:reco_current}.
 
The ratio of the average size of data errors over the total variation of the function is a probe of how faithful the reconstruction is to the fine features of the data. We find that the smaller this ratio is, the more the reconstruction is faithful to data features, independent of priors  and with well-defined posteriors for the hyperparameters. For current data this ratio is much larger than for the mock Stage IV data. If we multiply the current RSD measurements by a factor, or equivalently reduce their covariance by the same factor squared, the reconstruction improves.

In conclusion, we find that the distribution and precision of current data produces GP reconstructions that are strongly dominated by the choices of priors, either the range of the hyperparameters or the mean. We tested whether a multi-task GP adding the few measurements of $f$ and $\sig$ ameliorates the situation. The low number of data points and their precision does not allow us to evade the prior dominance. This scenario is also a typical example of an over-fitting configuration and should not be given much significance.

\section{Conclusion}\label{sec:discussion}

In this paper we explored Gaussian processes as a crucial tool for `agnostic' probes of gravity via the growth of large-scale structure. In particular, we focused on how GP can reconstruct the growth functions, namely the growth rate $f(z)$, the rms of matter fluctuations $\sigma_8(z)$ and their product $(f\sigma_8)(z)$.

We generated simulated measurements for the growth functions using a nominal sensitivity consistent with upcoming Stage IV surveys, and applied the GP pipeline to assess its ability to reconstruct the functions. Crucially, we reconstructed these functions not only separately, using single-task GP, but also via the multi-task GR approach. In multi-task GP, the three functions are reconstructed simultaneously, with a single likelihood. The kernel for the joint reconstruction contains sub-kernels for each function but also cross-convolution kernels that encode the links between the reconstructions. 

A multi-task approach is more faithful than single-task, since a common likelihood allows us to use a data covariance matrix which describes fully the data, i.e. it can take into account the possible correlations in the three data sets. At a more fundamental level, multi-task GP also incorporates the inter-dependence of the growth functions -- all three are probes of the evolution of matter perturbations. Furthermore, one function is the product of the other two.

We showed how the sharing of information between the reconstructions of each growth function in the multi-task approach is not only necessary but also enhances their precision. By performing a Monte Carlo generation of a large number of mocks, we show statistically that, relative to the single-task, the multi-task approach:

\begin{itemize}
    \item 
fits the data better; 
    \item 
is more precise; 
\item 
is more sensitive to the probability distribution of the data and thus can show more deviations from the fiducial. 
\end{itemize}

We verified the robustness of our findings against the change of the assumptions defining GP reconstructions.

Robustness however applies only to the simulated data sets of a nominal future survey. Indeed, we found that when applying our pipeline to currently available data, the reconstruction is very sensitive to the assumptions on the GP settings: priors on the hyperparameter ranges and the mean choices have a significant impact on the final reconstruction. We found that using our baseline settings (a zero prior mean and a very extended allowed range for the hyperparameters), the GP reconstructs a flat function. In order to reconstruct functions that behave as expected, one would instead need to fine tune the hyperparameters. This behaviour is due to the noise of current data being too random.

Next-generation surveys will reach a sensitivity that will enable the robust reconstruction of the growth functions $f(z)$, $\sigma_8(z)$ and $(f\sigma_8)(z)$, using model-agnostic methods such as GP. High-precision measurements will allow us to avoid fine tuning the parameters of the reconstruction, using only limited assumptions. The data-driven model-agnostic reconstruction of the three growth functions is further improved by a multi-task approach that incorporates the inter-dependence of these functions and possible correlations between the three data sets. A multi-task GP avoids the theoretical bias inherent in a single-task approach, which neglects inter-dependence and correlations. 

The ability of the GP to distinguish between a `true' deviation from a given model or an artefact from the noise of the data must be assessed with care. We showed that the accuracy of the reconstructions is highly dependent on the characteristics of the mock data, the function reconstructed and the GP method. Detecting a possible departure from the standard model with a GP reconstruction will require a more refined GP modelling in the future.

\section*{CRediT authorship contribution statement}

\textbf{Louis Perenon}: Conceptualization, Methodology, Software, Formal analysis, Investigation, Writing - Original Draft, Writing - Review \& Editing, Visualization, Supervision, Project administration. \textbf{Matteo Martinelli}: Conceptualization, Methodology, Software, Validation, Formal analysis, Writing - Original Draft, Writing - Review \& Editing. \textbf{St\'ephane Ili\'c}: Methodology, Validation, Writing - Review \& Editing. \textbf{Roy Maartens}: Conceptualization, Methodology, Writing - Original Draft, Writing - Review \& Editing, Supervision. \textbf{Michelle Lochner}: Methodology, Validation, Writing - Review \& Editing. \textbf{Chris Clarkson}: Conceptualization, Validation, Writing - Review \& Editing.  

\section*{Declaration of competing interest}

The authors declare that they have no known competing financial interests or personal relationships that
could have appeared to influence the work reported
in this paper.

\section*{Acknowledgements}

We thank Sambatra Andrianomena, Mario Ballardini, Stefano Camera, Ed Elson, Julien Larena, Marco Raveri and Mario Santos for useful discussions. We thank the authors of \cite{Shi:2017qpr,Jullo:2019lgq} for making their data covariance matrix available to us. We acknowledge the Sciama High Performance Compute Cluster, which is supported by the ICG at the University of Portsmouth, and the Centre for High Performance Computing, South Africa (under the project Cosmology with Radio Telescopes, ASTRO-0945), for providing computational resources for this research project. LP and RM are supported by the South African Radio Astronomy Observatory and the National Research Foundation (Grant No. 75415). MM is supported by a fellowship (code LCF/BQ/PI19/11690015) from `la Caixa' Foundation (ID 100010434) and by the Spanish Agencia Estatal de Investigacion through the grant IFT Centro de Excelencia Severo Ochoa SEV-2016-0597. RM is also supported by the United Kingdom Science \& Technology Facilities Council (STFC) Consolidated Grant ST/S000550/1. CC is supported by STFC Consolidated Grant ST/P000592/1.

\bibliographystyle{elsarticle-num-names}
\bibliography{References}

\begin{thebibliography}{74}
\expandafter\ifx\csname natexlab\endcsname\relax\def\natexlab#1{#1}\fi
\providecommand{\url}[1]{\texttt{#1}}
\providecommand{\href}[2]{#2}
\providecommand{\path}[1]{#1}
\providecommand{\DOIprefix}{doi:}
\providecommand{\ArXivprefix}{arXiv:}
\providecommand{\URLprefix}{URL: }
\providecommand{\Pubmedprefix}{pmid:}
\providecommand{\doi}[1]{\href{http://dx.doi.org/#1}{\path{#1}}}
\providecommand{\Pubmed}[1]{\href{pmid:#1}{\path{#1}}}
\providecommand{\bibinfo}[2]{#2}
\ifx\xfnm\relax \def\xfnm[#1]{\unskip,\space#1}\fi
\bibitem[{Blanchard et~al.(2020)}]{Blanchard:2019oqi}
\bibinfo{author}{A.~Blanchard}, et~al. (\bibinfo{collaboration}{Euclid}),
\newblock \bibinfo{title}{{Euclid preparation: VII. Forecast validation for
  Euclid cosmological probes}},
\newblock \bibinfo{journal}{Astron. Astrophys.} \bibinfo{volume}{642}
  (\bibinfo{year}{2020}) \bibinfo{pages}{A191}.
  \DOIprefix\doi{10.1051/0004-6361/202038071}.
  \href{http://arxiv.org/abs/1910.09273}{{\tt arXiv:1910.09273}}.
\bibitem[{Bacon et~al.(2020)}]{Bacon:2018dui}
\bibinfo{author}{D.~J. Bacon}, et~al. (\bibinfo{collaboration}{SKA}),
\newblock \bibinfo{title}{{Cosmology with Phase 1 of the Square Kilometre
  Array: Red Book 2018: Technical specifications and performance forecasts}},
\newblock \bibinfo{journal}{Publ. Astron. Soc. Austral.} \bibinfo{volume}{37}
  (\bibinfo{year}{2020}) \bibinfo{pages}{e007}.
  \DOIprefix\doi{10.1017/pasa.2019.51}.
  \href{http://arxiv.org/abs/1811.02743}{{\tt arXiv:1811.02743}}.
\bibitem[{Levi et~al.(2019)}]{Levi:2019ggs}
\bibinfo{author}{M.~E. Levi}, et~al. (\bibinfo{collaboration}{DESI}),
\newblock \bibinfo{title}{{The Dark Energy Spectroscopic Instrument (DESI)}}
  (\bibinfo{year}{2019}). \href{http://arxiv.org/abs/1907.10688}{{\tt
  arXiv:1907.10688}}.
\bibitem[{Dor\'e et~al.(2019)}]{Dore:2019pld}
\bibinfo{author}{O.~Dor\'e}, et~al.,
\newblock \bibinfo{title}{{WFIRST: The Essential Cosmology Space Observatory
  for the Coming Decade}}  (\bibinfo{year}{2019}).
  \href{http://arxiv.org/abs/1904.01174}{{\tt arXiv:1904.01174}}.
\bibitem[{Capozziello and De~Laurentis(2011)}]{Capozziello:2011et}
\bibinfo{author}{S.~Capozziello}, \bibinfo{author}{M.~De~Laurentis},
\newblock \bibinfo{title}{{Extended Theories of Gravity}},
\newblock \bibinfo{journal}{Phys. Rept.} \bibinfo{volume}{509}
  (\bibinfo{year}{2011}) \bibinfo{pages}{167--321}.
  \DOIprefix\doi{10.1016/j.physrep.2011.09.003}.
  \href{http://arxiv.org/abs/1108.6266}{{\tt arXiv:1108.6266}}.
\bibitem[{Clifton et~al.(2012)Clifton, Ferreira, Padilla, and
  Skordis}]{Clifton:2011jh}
\bibinfo{author}{T.~Clifton}, \bibinfo{author}{P.~G. Ferreira},
  \bibinfo{author}{A.~Padilla}, \bibinfo{author}{C.~Skordis},
\newblock \bibinfo{title}{{Modified Gravity and Cosmology}},
\newblock \bibinfo{journal}{Phys. Rept.} \bibinfo{volume}{513}
  (\bibinfo{year}{2012}) \bibinfo{pages}{1--189}.
  \DOIprefix\doi{10.1016/j.physrep.2012.01.001}.
  \href{http://arxiv.org/abs/1106.2476}{{\tt arXiv:1106.2476}}.
\bibitem[{Peirone et~al.(2019)Peirone, Benevento, Frusciante, and
  Tsujikawa}]{Peirone:2019aua}
\bibinfo{author}{S.~Peirone}, \bibinfo{author}{G.~Benevento},
  \bibinfo{author}{N.~Frusciante}, \bibinfo{author}{S.~Tsujikawa},
\newblock \bibinfo{title}{{Cosmological data favor Galileon ghost condensate
  over $\Lambda$CDM}},
\newblock \bibinfo{journal}{Phys. Rev. D} \bibinfo{volume}{100}
  (\bibinfo{year}{2019}) \bibinfo{pages}{063540}.
  \DOIprefix\doi{10.1103/PhysRevD.100.063540}.
  \href{http://arxiv.org/abs/1905.05166}{{\tt arXiv:1905.05166}}.
\bibitem[{Sol\`a~Peracaula et~al.(2019)Sol\`a~Peracaula, Gomez-Valent,
  de~Cruz~P\'erez, and Moreno-Pulido}]{Sola:2019jek}
\bibinfo{author}{J.~Sol\`a~Peracaula}, \bibinfo{author}{A.~Gomez-Valent},
  \bibinfo{author}{J.~de~Cruz~P\'erez}, \bibinfo{author}{C.~Moreno-Pulido},
\newblock \bibinfo{title}{{Brans\textendash{}Dicke Gravity with a Cosmological
  Constant Smoothes Out $\Lambda$CDM Tensions}},
\newblock \bibinfo{journal}{Astrophys. J. Lett.} \bibinfo{volume}{886}
  (\bibinfo{year}{2019}) \bibinfo{pages}{L6}.
  \DOIprefix\doi{10.3847/2041-8213/ab53e9}.
  \href{http://arxiv.org/abs/1909.02554}{{\tt arXiv:1909.02554}}.
\bibitem[{Aoki et~al.(2020)Aoki, De~Felice, Mukohyama, Noui, Oliosi, and
  Pookkillath}]{Aoki:2020oqc}
\bibinfo{author}{K.~Aoki}, \bibinfo{author}{A.~De~Felice},
  \bibinfo{author}{S.~Mukohyama}, \bibinfo{author}{K.~Noui},
  \bibinfo{author}{M.~Oliosi}, \bibinfo{author}{M.~C. Pookkillath},
\newblock \bibinfo{title}{{Minimally modified gravity fitting Planck data
  better than $\Lambda $CDM}},
\newblock \bibinfo{journal}{Eur. Phys. J. C} \bibinfo{volume}{80}
  (\bibinfo{year}{2020}) \bibinfo{pages}{708}.
  \DOIprefix\doi{10.1140/epjc/s10052-020-8291-1}.
  \href{http://arxiv.org/abs/2005.13972}{{\tt arXiv:2005.13972}}.
\bibitem[{Di~Valentino et~al.(2020)Di~Valentino, Melchiorri, Mena, Pan, and
  Yang}]{DiValentino:2020kpf}
\bibinfo{author}{E.~Di~Valentino}, \bibinfo{author}{A.~Melchiorri},
  \bibinfo{author}{O.~Mena}, \bibinfo{author}{S.~Pan},
  \bibinfo{author}{W.~Yang},
\newblock \bibinfo{title}{{Interacting Dark Energy in a closed universe}}
  (\bibinfo{year}{2020}). \DOIprefix\doi{10.1093/mnrasl/slaa207}.
  \href{http://arxiv.org/abs/2011.00283}{{\tt arXiv:2011.00283}}.
\bibitem[{Rasmussen and Williams(2006)}]{Rasmussen}
\bibinfo{author}{C.~Rasmussen}, \bibinfo{author}{C.~Williams},
  \bibinfo{title}{Gaussian Processes for Machine Learning}, Adaptive
  Computation and Machine Learning, \bibinfo{publisher}{MIT Press},
  \bibinfo{address}{Cambridge, MA, USA}, \bibinfo{year}{2006}.
\bibitem[{Holsclaw et~al.(2010{\natexlab{a}})Holsclaw, Alam, Sanso, Lee,
  Heitmann, Habib, and Higdon}]{Holsclaw:2010nb}
\bibinfo{author}{T.~Holsclaw}, \bibinfo{author}{U.~Alam},
  \bibinfo{author}{B.~Sanso}, \bibinfo{author}{H.~Lee},
  \bibinfo{author}{K.~Heitmann}, \bibinfo{author}{S.~Habib},
  \bibinfo{author}{D.~Higdon},
\newblock \bibinfo{title}{{Nonparametric Reconstruction of the Dark Energy
  Equation of State}},
\newblock \bibinfo{journal}{Phys. Rev. D} \bibinfo{volume}{82}
  (\bibinfo{year}{2010}{\natexlab{a}}) \bibinfo{pages}{103502}.
  \DOIprefix\doi{10.1103/PhysRevD.82.103502}.
  \href{http://arxiv.org/abs/1009.5443}{{\tt arXiv:1009.5443}}.
\bibitem[{Holsclaw et~al.(2010{\natexlab{b}})Holsclaw, Alam, Sanso, Lee,
  Heitmann, Habib, and Higdon}]{Holsclaw:2010sk}
\bibinfo{author}{T.~Holsclaw}, \bibinfo{author}{U.~Alam},
  \bibinfo{author}{B.~Sanso}, \bibinfo{author}{H.~Lee},
  \bibinfo{author}{K.~Heitmann}, \bibinfo{author}{S.~Habib},
  \bibinfo{author}{D.~Higdon},
\newblock \bibinfo{title}{{Nonparametric Dark Energy Reconstruction from
  Supernova Data}},
\newblock \bibinfo{journal}{Phys. Rev. Lett.} \bibinfo{volume}{105}
  (\bibinfo{year}{2010}{\natexlab{b}}) \bibinfo{pages}{241302}.
  \DOIprefix\doi{10.1103/PhysRevLett.105.241302}.
  \href{http://arxiv.org/abs/1011.3079}{{\tt arXiv:1011.3079}}.
\bibitem[{Holsclaw et~al.(2011)Holsclaw, Alam, Sanso, Lee, Heitmann, Habib, and
  Higdon}]{Holsclaw:2011wi}
\bibinfo{author}{T.~Holsclaw}, \bibinfo{author}{U.~Alam},
  \bibinfo{author}{B.~Sanso}, \bibinfo{author}{H.~Lee},
  \bibinfo{author}{K.~Heitmann}, \bibinfo{author}{S.~Habib},
  \bibinfo{author}{D.~Higdon},
\newblock \bibinfo{title}{{Nonparametric Reconstruction of the Dark Energy
  Equation of State from Diverse Data Sets}},
\newblock \bibinfo{journal}{Phys. Rev. D} \bibinfo{volume}{84}
  (\bibinfo{year}{2011}) \bibinfo{pages}{083501}.
  \DOIprefix\doi{10.1103/PhysRevD.84.083501}.
  \href{http://arxiv.org/abs/1104.2041}{{\tt arXiv:1104.2041}}.
\bibitem[{Shafieloo et~al.(2012)Shafieloo, Kim, and Linder}]{Shafieloo:2012ht}
\bibinfo{author}{A.~Shafieloo}, \bibinfo{author}{A.~G. Kim},
  \bibinfo{author}{E.~V. Linder},
\newblock \bibinfo{title}{{Gaussian Process Cosmography}},
\newblock \bibinfo{journal}{Phys. Rev. D} \bibinfo{volume}{85}
  (\bibinfo{year}{2012}) \bibinfo{pages}{123530}.
  \DOIprefix\doi{10.1103/PhysRevD.85.123530}.
  \href{http://arxiv.org/abs/1204.2272}{{\tt arXiv:1204.2272}}.
\bibitem[{Seikel et~al.(2012)Seikel, Clarkson, and Smith}]{Seikel:2012uu}
\bibinfo{author}{M.~Seikel}, \bibinfo{author}{C.~Clarkson},
  \bibinfo{author}{M.~Smith},
\newblock \bibinfo{title}{{Reconstruction of dark energy and expansion dynamics
  using Gaussian processes}},
\newblock \bibinfo{journal}{JCAP} \bibinfo{volume}{06} (\bibinfo{year}{2012})
  \bibinfo{pages}{036}. \DOIprefix\doi{10.1088/1475-7516/2012/06/036}.
  \href{http://arxiv.org/abs/1204.2832}{{\tt arXiv:1204.2832}}.
\bibitem[{Yahya et~al.(2014)Yahya, Seikel, Clarkson, Maartens, and
  Smith}]{Yahya:2013xma}
\bibinfo{author}{S.~Yahya}, \bibinfo{author}{M.~Seikel},
  \bibinfo{author}{C.~Clarkson}, \bibinfo{author}{R.~Maartens},
  \bibinfo{author}{M.~Smith},
\newblock \bibinfo{title}{{Null tests of the cosmological constant using
  supernovae}},
\newblock \bibinfo{journal}{Phys. Rev. D} \bibinfo{volume}{89}
  (\bibinfo{year}{2014}) \bibinfo{pages}{023503}.
  \DOIprefix\doi{10.1103/PhysRevD.89.023503}.
  \href{http://arxiv.org/abs/1308.4099}{{\tt arXiv:1308.4099}}.
\bibitem[{Bester et~al.(2014)Bester, Larena, van~der Walt, and
  Bishop}]{Bester:2013fya}
\bibinfo{author}{H.~L. Bester}, \bibinfo{author}{J.~Larena},
  \bibinfo{author}{P.~J. van~der Walt}, \bibinfo{author}{N.~T. Bishop},
\newblock \bibinfo{title}{{What's Inside the Cone? Numerically reconstructing
  the metric from observations}},
\newblock \bibinfo{journal}{JCAP} \bibinfo{volume}{02} (\bibinfo{year}{2014})
  \bibinfo{pages}{009}. \DOIprefix\doi{10.1088/1475-7516/2014/02/009}.
  \href{http://arxiv.org/abs/1312.1081}{{\tt arXiv:1312.1081}}.
\bibitem[{Busti et~al.(2014)Busti, Clarkson, and Seikel}]{Busti:2014dua}
\bibinfo{author}{V.~C. Busti}, \bibinfo{author}{C.~Clarkson},
  \bibinfo{author}{M.~Seikel},
\newblock \bibinfo{title}{{Evidence for a Lower Value for $H_0$ from Cosmic
  Chronometers Data?}},
\newblock \bibinfo{journal}{Mon. Not. Roy. Astron. Soc.} \bibinfo{volume}{441}
  (\bibinfo{year}{2014}) \bibinfo{pages}{11}.
  \DOIprefix\doi{10.1093/mnrasl/slu035}.
  \href{http://arxiv.org/abs/1402.5429}{{\tt arXiv:1402.5429}}.
\bibitem[{Bester et~al.(2015)Bester, Larena, and Bishop}]{Bester:2015gla}
\bibinfo{author}{H.~L. Bester}, \bibinfo{author}{J.~Larena},
  \bibinfo{author}{N.~T. Bishop},
\newblock \bibinfo{title}{{Towards the geometry of the universe from data}},
\newblock \bibinfo{journal}{Mon. Not. Roy. Astron. Soc.} \bibinfo{volume}{453}
  (\bibinfo{year}{2015}) \bibinfo{pages}{2364--2377}.
  \DOIprefix\doi{10.1093/mnras/stv1672}.
  \href{http://arxiv.org/abs/1506.01591}{{\tt arXiv:1506.01591}}.
\bibitem[{Bester et~al.(2017)Bester, Larena, and Bishop}]{Bester:2016fbs}
\bibinfo{author}{H.~L. Bester}, \bibinfo{author}{J.~Larena},
  \bibinfo{author}{N.~T. Bishop},
\newblock \bibinfo{title}{{Numerically reconstructing the geometry of the
  Universe from data}},
\newblock in: \bibinfo{booktitle}{{14th Marcel Grossmann Meeting on Recent
  Developments in Theoretical and Experimental General Relativity,
  Astrophysics, and Relativistic Field Theories}}, volume~\bibinfo{volume}{3},
  \bibinfo{year}{2017}, pp. \bibinfo{pages}{2284--2289}.
  \DOIprefix\doi{10.1142/9789813226609_0263}.
  \href{http://arxiv.org/abs/1601.07362}{{\tt arXiv:1601.07362}}.
\bibitem[{Joudaki et~al.(2018)Joudaki, Kaplinghat, Keeley, and
  Kirkby}]{Joudaki:2017zhq}
\bibinfo{author}{S.~Joudaki}, \bibinfo{author}{M.~Kaplinghat},
  \bibinfo{author}{R.~Keeley}, \bibinfo{author}{D.~Kirkby},
\newblock \bibinfo{title}{{Model independent inference of the expansion history
  and implications for the growth of structure}},
\newblock \bibinfo{journal}{Phys. Rev. D} \bibinfo{volume}{97}
  (\bibinfo{year}{2018}) \bibinfo{pages}{123501}.
  \DOIprefix\doi{10.1103/PhysRevD.97.123501}.
  \href{http://arxiv.org/abs/1710.04236}{{\tt arXiv:1710.04236}}.
\bibitem[{G\'omez-Valent and Amendola(2018)}]{Gomez-Valent:2018hwc}
\bibinfo{author}{A.~G\'omez-Valent}, \bibinfo{author}{L.~Amendola},
\newblock \bibinfo{title}{{$H_0$ from cosmic chronometers and Type Ia
  supernovae, with Gaussian Processes and the novel Weighted Polynomial
  Regression method}},
\newblock \bibinfo{journal}{JCAP} \bibinfo{volume}{04} (\bibinfo{year}{2018})
  \bibinfo{pages}{051}. \DOIprefix\doi{10.1088/1475-7516/2018/04/051}.
  \href{http://arxiv.org/abs/1802.01505}{{\tt arXiv:1802.01505}}.
\bibitem[{Haridasu et~al.(2018)Haridasu, Lukovi\'c, Moresco, and
  Vittorio}]{Haridasu:2018gqm}
\bibinfo{author}{B.~S. Haridasu}, \bibinfo{author}{V.~V. Lukovi\'c},
  \bibinfo{author}{M.~Moresco}, \bibinfo{author}{N.~Vittorio},
\newblock \bibinfo{title}{{An improved model-independent assessment of the
  late-time cosmic expansion}},
\newblock \bibinfo{journal}{JCAP} \bibinfo{volume}{10} (\bibinfo{year}{2018})
  \bibinfo{pages}{015}. \DOIprefix\doi{10.1088/1475-7516/2018/10/015}.
  \href{http://arxiv.org/abs/1805.03595}{{\tt arXiv:1805.03595}}.
\bibitem[{Gerardi et~al.(2019)Gerardi, Martinelli, and
  Silvestri}]{Gerardi:2019obr}
\bibinfo{author}{F.~Gerardi}, \bibinfo{author}{M.~Martinelli},
  \bibinfo{author}{A.~Silvestri},
\newblock \bibinfo{title}{{Reconstruction of the Dark Energy equation of state
  from latest data: the impact of theoretical priors}},
\newblock \bibinfo{journal}{JCAP} \bibinfo{volume}{07} (\bibinfo{year}{2019})
  \bibinfo{pages}{042}. \DOIprefix\doi{10.1088/1475-7516/2019/07/042}.
  \href{http://arxiv.org/abs/1902.09423}{{\tt arXiv:1902.09423}}.
\bibitem[{Keeley et~al.(2020)Keeley, Shafieloo, L'Huillier, and
  Linder}]{Keeley:2019hmw}
\bibinfo{author}{R.~E. Keeley}, \bibinfo{author}{A.~Shafieloo},
  \bibinfo{author}{B.~L'Huillier}, \bibinfo{author}{E.~V. Linder},
\newblock \bibinfo{title}{{Debiasing Cosmic Gravitational Wave Sirens}},
\newblock \bibinfo{journal}{Mon. Not. Roy. Astron. Soc.} \bibinfo{volume}{491}
  (\bibinfo{year}{2020}) \bibinfo{pages}{3983--3989}.
  \DOIprefix\doi{10.1093/mnras/stz3304}.
  \href{http://arxiv.org/abs/1905.10216}{{\tt arXiv:1905.10216}}.
\bibitem[{Bengaly et~al.(2020)Bengaly, Clarkson, and
  Maartens}]{Bengaly:2019oxx}
\bibinfo{author}{C.~A.~P. Bengaly}, \bibinfo{author}{C.~Clarkson},
  \bibinfo{author}{R.~Maartens},
\newblock \bibinfo{title}{{The Hubble constant tension with next-generation
  galaxy surveys}},
\newblock \bibinfo{journal}{JCAP} \bibinfo{volume}{05} (\bibinfo{year}{2020})
  \bibinfo{pages}{053}. \DOIprefix\doi{10.1088/1475-7516/2020/05/053}.
  \href{http://arxiv.org/abs/1908.04619}{{\tt arXiv:1908.04619}}.
\bibitem[{Bengaly(2020)}]{Bengaly:2019ibu}
\bibinfo{author}{C.~A.~P. Bengaly},
\newblock \bibinfo{title}{{Evidence for cosmic acceleration with
  next-generation surveys: A model-independent approach}},
\newblock \bibinfo{journal}{Mon. Not. Roy. Astron. Soc.} \bibinfo{volume}{499}
  (\bibinfo{year}{2020}) \bibinfo{pages}{L6--L10}.
  \DOIprefix\doi{10.1093/mnrasl/slaa040}.
  \href{http://arxiv.org/abs/1912.05528}{{\tt arXiv:1912.05528}}.
\bibitem[{Aljaf et~al.(2020)Aljaf, Gregoris, and Khurshudyan}]{Aljaf:2020eqh}
\bibinfo{author}{M.~Aljaf}, \bibinfo{author}{D.~Gregoris},
  \bibinfo{author}{M.~Khurshudyan},
\newblock \bibinfo{title}{{Constraints on interacting dark energy models
  through cosmic chronometers and Gaussian process}}  (\bibinfo{year}{2020}).
  \href{http://arxiv.org/abs/2005.01891}{{\tt arXiv:2005.01891}}.
\bibitem[{Colg\'ain and Sheikh-Jabbari(2021)}]{Colgain:2021ngq}
\bibinfo{author}{E.~O. Colg\'ain}, \bibinfo{author}{M.~M. Sheikh-Jabbari},
\newblock \bibinfo{title}{{On model independent cosmic determinations of
  $H_0$}}  (\bibinfo{year}{2021}). \href{http://arxiv.org/abs/2101.08565}{{\tt
  arXiv:2101.08565}}.
\bibitem[{Dhawan et~al.(2021)Dhawan, Alsing, and Vagnozzi}]{Dhawan:2021mel}
\bibinfo{author}{S.~Dhawan}, \bibinfo{author}{J.~Alsing},
  \bibinfo{author}{S.~Vagnozzi},
\newblock \bibinfo{title}{{Non-parametric spatial curvature inference using
  late-universe cosmological probes}}  (\bibinfo{year}{2021}).
  \href{http://arxiv.org/abs/2104.02485}{{\tt arXiv:2104.02485}}.
\bibitem[{Mukherjee and Mukherjee(2021)}]{Mukherjee:2021kcu}
\bibinfo{author}{P.~Mukherjee}, \bibinfo{author}{A.~Mukherjee},
\newblock \bibinfo{title}{{Assessment of the cosmic distance duality relation
  using Gaussian Process}}  (\bibinfo{year}{2021}).
  \href{http://arxiv.org/abs/2104.06066}{{\tt arXiv:2104.06066}}.
\bibitem[{{Velasquez-Toribio} and {Fabris}(2021)}]{2021arXiv210407356V}
\bibinfo{author}{A.~M. {Velasquez-Toribio}}, \bibinfo{author}{J.~C. {Fabris}},
\newblock \bibinfo{title}{{Constraints on Cosmographic Functions using Gaussian
  Processes}},
\newblock \bibinfo{journal}{arXiv e-prints}  (\bibinfo{year}{2021})
  \bibinfo{pages}{arXiv:2104.07356}.
  \href{http://arxiv.org/abs/2104.07356}{{\tt arXiv:2104.07356}}.
\bibitem[{Ca\~nas Herrera et~al.(2021)Ca\~nas Herrera, Contigiani, and
  Vardanyan}]{Canas-Herrera:2021qxs}
\bibinfo{author}{G.~Ca\~nas Herrera}, \bibinfo{author}{O.~Contigiani},
  \bibinfo{author}{V.~Vardanyan},
\newblock \bibinfo{title}{{Learning How to Surf: Reconstructing the Propagation
  and Origin of Gravitational Waves with Gaussian Processes}},
\newblock \bibinfo{journal}{Astrophys. J.} \bibinfo{volume}{918}
  (\bibinfo{year}{2021}) \bibinfo{pages}{20}.
  \DOIprefix\doi{10.3847/1538-4357/ac09e3}.
  \href{http://arxiv.org/abs/2105.04262}{{\tt arXiv:2105.04262}}.
\bibitem[{Liao et~al.(2019)Liao, Shafieloo, Keeley, and Linder}]{Liao:2019qoc}
\bibinfo{author}{K.~Liao}, \bibinfo{author}{A.~Shafieloo},
  \bibinfo{author}{R.~E. Keeley}, \bibinfo{author}{E.~V. Linder},
\newblock \bibinfo{title}{{A model-independent determination of the Hubble
  constant from lensed quasars and supernovae using Gaussian process
  regression}},
\newblock \bibinfo{journal}{Astrophys. J. Lett.} \bibinfo{volume}{886}
  (\bibinfo{year}{2019}) \bibinfo{pages}{L23}.
  \DOIprefix\doi{10.3847/2041-8213/ab5308}.
  \href{http://arxiv.org/abs/1908.04967}{{\tt arXiv:1908.04967}}.
\bibitem[{Liu et~al.(2019)Liu, Cao, Zhang, Geng, Liu, Ji, and
  Zhu}]{Liu:2019ddm}
\bibinfo{author}{T.~Liu}, \bibinfo{author}{S.~Cao}, \bibinfo{author}{J.~Zhang},
  \bibinfo{author}{S.~Geng}, \bibinfo{author}{Y.~Liu}, \bibinfo{author}{X.~Ji},
  \bibinfo{author}{Z.-H. Zhu},
\newblock \bibinfo{title}{{Implications from simulated strong gravitational
  lensing systems: constraining cosmological parameters using Gaussian
  Processes}},
\newblock \bibinfo{journal}{Astrophys. J.} \bibinfo{volume}{886}
  (\bibinfo{year}{2019}) \bibinfo{pages}{94}.
  \DOIprefix\doi{10.3847/1538-4357/ab4bc3}.
  \href{http://arxiv.org/abs/1910.02592}{{\tt arXiv:1910.02592}}.
\bibitem[{Pandey et~al.(2020)Pandey, Raveri, and Jain}]{Pandey:2019yic}
\bibinfo{author}{S.~Pandey}, \bibinfo{author}{M.~Raveri},
  \bibinfo{author}{B.~Jain},
\newblock \bibinfo{title}{{Model independent comparison of supernova and strong
  lensing cosmography: Implications for the Hubble constant tension}},
\newblock \bibinfo{journal}{Phys. Rev. D} \bibinfo{volume}{102}
  (\bibinfo{year}{2020}) \bibinfo{pages}{023505}.
  \DOIprefix\doi{10.1103/PhysRevD.102.023505}.
  \href{http://arxiv.org/abs/1912.04325}{{\tt arXiv:1912.04325}}.
\bibitem[{Liao et~al.(2020)Liao, Shafieloo, Keeley, and Linder}]{Liao:2020zko}
\bibinfo{author}{K.~Liao}, \bibinfo{author}{A.~Shafieloo},
  \bibinfo{author}{R.~E. Keeley}, \bibinfo{author}{E.~V. Linder},
\newblock \bibinfo{title}{{Determining Model-independent H 0 and Consistency
  Tests}},
\newblock \bibinfo{journal}{Astrophys. J. Lett.} \bibinfo{volume}{895}
  (\bibinfo{year}{2020}) \bibinfo{pages}{L29}.
  \DOIprefix\doi{10.3847/2041-8213/ab8dbb}.
  \href{http://arxiv.org/abs/2002.10605}{{\tt arXiv:2002.10605}}.
\bibitem[{Renzi and Silvestri(2020)}]{Renzi:2020fnx}
\bibinfo{author}{F.~Renzi}, \bibinfo{author}{A.~Silvestri},
\newblock \bibinfo{title}{{A look at the Hubble speed from first principles}}
  (\bibinfo{year}{2020}). \href{http://arxiv.org/abs/2011.10559}{{\tt
  arXiv:2011.10559}}.
\bibitem[{Bengaly et~al.(2020)Bengaly, Clarkson, Kunz, and
  Maartens}]{Bengaly:2020neu}
\bibinfo{author}{C.~A.~P. Bengaly}, \bibinfo{author}{C.~Clarkson},
  \bibinfo{author}{M.~Kunz}, \bibinfo{author}{R.~Maartens},
\newblock \bibinfo{title}{{Null tests of the concordance model in the era of
  Euclid and the SKA}}  (\bibinfo{year}{2020}).
  \href{http://arxiv.org/abs/2007.04879}{{\tt arXiv:2007.04879}}.
\bibitem[{Zhang et~al.(2021)Zhang, Jiao, and Zhang}]{Zhang:2021tmg}
\bibinfo{author}{J.-C. Zhang}, \bibinfo{author}{K.~Jiao},
  \bibinfo{author}{T.-J. Zhang},
\newblock \bibinfo{title}{{Model-independent measurement of the Hubble Constant
  and the absolute magnitude of Type Ia Supernovae}}  (\bibinfo{year}{2021}).
  \href{http://arxiv.org/abs/2101.05897}{{\tt arXiv:2101.05897}}.
\bibitem[{{Almosallam} et~al.(2016){Almosallam}, {Jarvis}, and
  {Roberts}}]{2016MNRAS.462..726A}
\bibinfo{author}{I.~A. {Almosallam}}, \bibinfo{author}{M.~J. {Jarvis}},
  \bibinfo{author}{S.~J. {Roberts}},
\newblock \bibinfo{title}{{GPZ: non-stationary sparse Gaussian processes for
  heteroscedastic uncertainty estimation in photometric redshifts}},
\newblock \bibinfo{journal}{Mon. Not. Roy. Astron. Soc.} \bibinfo{volume}{462}
  (\bibinfo{year}{2016}) \bibinfo{pages}{726--739}.
  \DOIprefix\doi{10.1093/mnras/stw1618}.
  \href{http://arxiv.org/abs/1604.03593}{{\tt arXiv:1604.03593}}.
\bibitem[{Mootoovaloo et~al.(2020)Mootoovaloo, Heavens, Jaffe, and
  Leclercq}]{Mootoovaloo:2020ott}
\bibinfo{author}{A.~Mootoovaloo}, \bibinfo{author}{A.~F. Heavens},
  \bibinfo{author}{A.~H. Jaffe}, \bibinfo{author}{F.~Leclercq},
\newblock \bibinfo{title}{{Parameter Inference for Weak Lensing using Gaussian
  Processes and MOPED}},
\newblock \bibinfo{journal}{Mon. Not. Roy. Astron. Soc.} \bibinfo{volume}{497}
  (\bibinfo{year}{2020}) \bibinfo{pages}{2213--2226}.
  \DOIprefix\doi{10.1093/mnras/staa2102}.
  \href{http://arxiv.org/abs/2005.06551}{{\tt arXiv:2005.06551}}.
\bibitem[{Pinho et~al.(2018)Pinho, Casas, and Amendola}]{Pinho:2018unz}
\bibinfo{author}{A.~M. Pinho}, \bibinfo{author}{S.~Casas},
  \bibinfo{author}{L.~Amendola},
\newblock \bibinfo{title}{{Model-independent reconstruction of the linear
  anisotropic stress $\eta$}},
\newblock \bibinfo{journal}{JCAP} \bibinfo{volume}{11} (\bibinfo{year}{2018})
  \bibinfo{pages}{027}. \DOIprefix\doi{10.1088/1475-7516/2018/11/027}.
  \href{http://arxiv.org/abs/1805.00027}{{\tt arXiv:1805.00027}}.
\bibitem[{Zhang and Li(2018)}]{Zhang:2018gjb}
\bibinfo{author}{M.-J. Zhang}, \bibinfo{author}{H.~Li},
\newblock \bibinfo{title}{{Gaussian processes reconstruction of dark energy
  from observational data}},
\newblock \bibinfo{journal}{Eur. Phys. J. C} \bibinfo{volume}{78}
  (\bibinfo{year}{2018}) \bibinfo{pages}{460}.
  \DOIprefix\doi{10.1140/epjc/s10052-018-5953-3}.
  \href{http://arxiv.org/abs/1806.02981}{{\tt arXiv:1806.02981}}.
\bibitem[{Yin and Wei(2019)}]{Yin:2018mvu}
\bibinfo{author}{Z.-Y. Yin}, \bibinfo{author}{H.~Wei},
\newblock \bibinfo{title}{{Non-parametric Reconstruction of Growth Index via
  Gaussian Processes}},
\newblock \bibinfo{journal}{Sci. China Phys. Mech. Astron.}
  \bibinfo{volume}{62} (\bibinfo{year}{2019}) \bibinfo{pages}{999811}.
  \DOIprefix\doi{10.1007/s11433-019-9373-0}.
  \href{http://arxiv.org/abs/1808.00377}{{\tt arXiv:1808.00377}}.
\bibitem[{Li et~al.(2021)Li, Du, Zhou, Zhang, and Xu}]{Li:2019nux}
\bibinfo{author}{E.-K. Li}, \bibinfo{author}{M.~Du}, \bibinfo{author}{Z.-H.
  Zhou}, \bibinfo{author}{H.~Zhang}, \bibinfo{author}{L.~Xu},
\newblock \bibinfo{title}{{Testing the effect of $H_0$ on $f\sigma_8$ tension
  using a Gaussian process method}},
\newblock \bibinfo{journal}{Mon. Not. Roy. Astron. Soc.} \bibinfo{volume}{501}
  (\bibinfo{year}{2021}) \bibinfo{pages}{4452--4463}.
  \DOIprefix\doi{10.1093/mnras/staa3894}.
  \href{http://arxiv.org/abs/1911.12076}{{\tt arXiv:1911.12076}}.
\bibitem[{Benisty(2021)}]{Benisty:2020kdt}
\bibinfo{author}{D.~Benisty},
\newblock \bibinfo{title}{{Quantifying the $S_8$ tension with the Redshift
  Space Distortion data set}},
\newblock \bibinfo{journal}{Phys. Dark Univ.} \bibinfo{volume}{1}
  (\bibinfo{year}{2021}) \bibinfo{pages}{100766}.
  \DOIprefix\doi{10.1016/j.dark.2020.100766}.
  \href{http://arxiv.org/abs/2005.03751}{{\tt arXiv:2005.03751}}.
\bibitem[{Caruana(1998)}]{Caruana1998}
\bibinfo{author}{R.~Caruana}, \bibinfo{title}{Multitask Learning},
  \bibinfo{publisher}{Springer US}, \bibinfo{address}{Boston, MA},
  \bibinfo{year}{1998}, pp. \bibinfo{pages}{95--133}. \URLprefix
  \url{https://doi.org/10.1007/978-1-4615-5529-2_5}.
  \DOIprefix\doi{10.1007/978-1-4615-5529-2_5}.
\bibitem[{Bonilla et~al.(2007)Bonilla, Agakov, and
  Williams}]{866abfcac5f9481d97628a546255878b}
\bibinfo{author}{E.~Bonilla}, \bibinfo{author}{F.~Agakov},
  \bibinfo{author}{C.~Williams},
\newblock \bibinfo{title}{Kernel multi-task learning using task-specific
  features},
\newblock in: \bibinfo{editor}{M.~Meila}, \bibinfo{editor}{X.~Shen} (Eds.),
  \bibinfo{booktitle}{Proceedings of the Eleventh International Conference on
  Artificial Intelligence and Statistics (AISTATS 2007)},
  volume~\bibinfo{volume}{2}, \bibinfo{publisher}{Journal of Machine Learning
  Research: Workshop and Conference Proceedings}, \bibinfo{year}{2007}, pp.
  \bibinfo{pages}{43--50}.
\bibitem[{Bonilla et~al.(2008)Bonilla, Chai, and
  Williams}]{21ae58e5ad934d7494d64e5bfd4a1a52}
\bibinfo{author}{E.~Bonilla}, \bibinfo{author}{K.~Chai},
  \bibinfo{author}{C.~Williams},
\newblock \bibinfo{title}{Multi-task gaussian process prediction},
\newblock in: \bibinfo{booktitle}{Advances in Neural Information Processing
  Systems 20}, \bibinfo{publisher}{NIPS Foundation}, \bibinfo{year}{2008}, pp.
  \bibinfo{pages}{153--160}.
\bibitem[{Melkumyan and Ramos(2011)}]{Melkumyan}
\bibinfo{author}{A.~Melkumyan}, \bibinfo{author}{F.~Ramos},
\newblock \bibinfo{title}{Multi-kernel gaussian processes.},
\newblock \bibinfo{journal}{IJCAI International Joint Conference on Artificial
  Intelligence}  (\bibinfo{year}{2011}) \bibinfo{pages}{1408--1413}.
  \DOIprefix\doi{10.5591/978-1-57735-516-8/IJCAI11-238}.
\bibitem[{Vasudevan(2012)}]{Vasudevan}
\bibinfo{author}{S.~Vasudevan},
\newblock \bibinfo{title}{Data fusion with gaussian processes},
\newblock \bibinfo{journal}{Robotics and Autonomous Systems}
  \bibinfo{volume}{60} (\bibinfo{year}{2012}) \bibinfo{pages}{1528–1544}.
  \DOIprefix\doi{10.1016/j.robot.2012.08.006}.
\bibitem[{{Sargent} and {Turner}(1977)}]{1977ApJ...212L...3S}
\bibinfo{author}{W.~L.~W. {Sargent}}, \bibinfo{author}{E.~L. {Turner}},
\newblock \bibinfo{title}{{A statistical method for determining the
  cosmological density parameter from the redshifts of a complete sample of
  galaxies.}},
\newblock \bibinfo{journal}{Astrophysical Journal} \bibinfo{volume}{212}
  (\bibinfo{year}{1977}) \bibinfo{pages}{L3--L7}.
  \DOIprefix\doi{10.1086/182362}.
\bibitem[{{Kaiser}(1987)}]{Kaiser:1987qv}
\bibinfo{author}{N.~{Kaiser}},
\newblock \bibinfo{title}{{Clustering in real space and in redshift space}},
\newblock \bibinfo{journal}{Mon. Not. Roy. Astron. Soc.} \bibinfo{volume}{227}
  (\bibinfo{year}{1987}) \bibinfo{pages}{1--21}.
  \DOIprefix\doi{10.1093/mnras/227.1.1}.
\bibitem[{Guzzo et~al.(2008)}]{Guzzo:2008ac}
\bibinfo{author}{L.~Guzzo}, et~al.,
\newblock \bibinfo{title}{{A test of the nature of cosmic acceleration using
  galaxy redshift distortions}},
\newblock \bibinfo{journal}{Nature} \bibinfo{volume}{451}
  (\bibinfo{year}{2008}) \bibinfo{pages}{541--545}.
  \DOIprefix\doi{10.1038/nature06555}.
  \href{http://arxiv.org/abs/0802.1944}{{\tt arXiv:0802.1944}}.
\bibitem[{Song and Percival(2009)}]{Song:2008qt}
\bibinfo{author}{Y.-S. Song}, \bibinfo{author}{W.~J. Percival},
\newblock \bibinfo{title}{{Reconstructing the history of structure formation
  using Redshift Distortions}},
\newblock \bibinfo{journal}{JCAP} \bibinfo{volume}{0910} (\bibinfo{year}{2009})
  \bibinfo{pages}{004}. \DOIprefix\doi{10.1088/1475-7516/2009/10/004}.
  \href{http://arxiv.org/abs/0807.0810}{{\tt arXiv:0807.0810}}.
\bibitem[{Percival and White(2009)}]{Percival:2008sh}
\bibinfo{author}{W.~J. Percival}, \bibinfo{author}{M.~White},
\newblock \bibinfo{title}{{Testing cosmological structure formation using
  redshift-space distortions}},
\newblock \bibinfo{journal}{Mon. Not. Roy. Astron. Soc.} \bibinfo{volume}{393}
  (\bibinfo{year}{2009}) \bibinfo{pages}{297}.
  \DOIprefix\doi{10.1111/j.1365-2966.2008.14211.x}.
  \href{http://arxiv.org/abs/0808.0003}{{\tt arXiv:0808.0003}}.
\bibitem[{Gil-Marín et~al.(2017)Gil-Marín, Percival, Verde, Brownstein,
  Chuang, Kitaura, Rodríguez-Torres, and Olmstead}]{Gil-Marin:2016wya}
\bibinfo{author}{H.~Gil-Marín}, \bibinfo{author}{W.~J. Percival},
  \bibinfo{author}{L.~Verde}, \bibinfo{author}{J.~R. Brownstein},
  \bibinfo{author}{C.-H. Chuang}, \bibinfo{author}{F.-S. Kitaura},
  \bibinfo{author}{S.~A. Rodríguez-Torres}, \bibinfo{author}{M.~D. Olmstead},
\newblock \bibinfo{title}{{The clustering of galaxies in the SDSS-III Baryon
  Oscillation Spectroscopic Survey: RSD measurement from the power spectrum and
  bispectrum of the DR12 BOSS galaxies}},
\newblock \bibinfo{journal}{Mon. Not. Roy. Astron. Soc.} \bibinfo{volume}{465}
  (\bibinfo{year}{2017}) \bibinfo{pages}{1757--1788}.
  \DOIprefix\doi{10.1093/mnras/stw2679}.
  \href{http://arxiv.org/abs/1606.00439}{{\tt arXiv:1606.00439}}.
\bibitem[{de~la Torre et~al.(2017)}]{delaTorre:2016rxm}
\bibinfo{author}{S.~de~la Torre}, et~al.,
\newblock \bibinfo{title}{{The VIMOS Public Extragalactic Redshift Survey
  (VIPERS). Gravity test from the combination of redshift-space distortions and
  galaxy-galaxy lensing at $0.5 < z < 1.2$}},
\newblock \bibinfo{journal}{Astron. Astrophys.} \bibinfo{volume}{608}
  (\bibinfo{year}{2017}) \bibinfo{pages}{A44}.
  \DOIprefix\doi{10.1051/0004-6361/201630276}.
  \href{http://arxiv.org/abs/1612.05647}{{\tt arXiv:1612.05647}}.
\bibitem[{Shi et~al.(2018)}]{Shi:2017qpr}
\bibinfo{author}{F.~Shi}, et~al.,
\newblock \bibinfo{title}{{Mapping the Real Space Distributions of Galaxies in
  SDSS DR7: II. Measuring the growth rate, clustering amplitude of matter and
  biases of galaxies at redshift $0.1$}},
\newblock \bibinfo{journal}{Astrophys. J.} \bibinfo{volume}{861}
  (\bibinfo{year}{2018}) \bibinfo{pages}{137}.
  \DOIprefix\doi{10.3847/1538-4357/aacb20}.
  \href{http://arxiv.org/abs/1712.04163}{{\tt arXiv:1712.04163}}.
\bibitem[{Jullo et~al.(2019)}]{Jullo:2019lgq}
\bibinfo{author}{E.~Jullo}, et~al.,
\newblock \bibinfo{title}{{Testing gravity with galaxy-galaxy lensing and
  redshift-space distortions using CFHT-Stripe 82, CFHTLenS and BOSS CMASS
  datasets}},
\newblock \bibinfo{journal}{Astron. Astrophys.} \bibinfo{volume}{627}
  (\bibinfo{year}{2019}) \bibinfo{pages}{A137}.
  \DOIprefix\doi{10.1051/0004-6361/201834629}.
  \href{http://arxiv.org/abs/1903.07160}{{\tt arXiv:1903.07160}}.
\bibitem[{Seikel and Clarkson(2013)}]{Seikel:2013fda}
\bibinfo{author}{M.~Seikel}, \bibinfo{author}{C.~Clarkson},
\newblock \bibinfo{title}{{Optimising Gaussian processes for reconstructing
  dark energy dynamics from supernovae}}  (\bibinfo{year}{2013}).
  \href{http://arxiv.org/abs/1311.6678}{{\tt arXiv:1311.6678}}.
\bibitem[{Aghanim et~al.(2020)}]{Aghanim:2018eyx}
\bibinfo{author}{N.~Aghanim}, et~al. (\bibinfo{collaboration}{Planck}),
\newblock \bibinfo{title}{{Planck 2018 results. VI. Cosmological parameters}},
\newblock \bibinfo{journal}{Astron. Astrophys.} \bibinfo{volume}{641}
  (\bibinfo{year}{2020}) \bibinfo{pages}{A6}.
  \DOIprefix\doi{10.1051/0004-6361/201833910}.
  \href{http://arxiv.org/abs/1807.06209}{{\tt arXiv:1807.06209}}.
\bibitem[{{Blas} et~al.(2011){Blas}, {Lesgourgues}, and
  {Tram}}]{2011JCAP...07..034B}
\bibinfo{author}{D.~{Blas}}, \bibinfo{author}{J.~{Lesgourgues}},
  \bibinfo{author}{T.~{Tram}},
\newblock \bibinfo{title}{{The Cosmic Linear Anisotropy Solving System (CLASS).
  Part II: Approximation schemes}},
\newblock \bibinfo{journal}{"JCAP"} \bibinfo{volume}{2011}
  (\bibinfo{year}{2011}) \bibinfo{pages}{034}.
  \DOIprefix\doi{10.1088/1475-7516/2011/07/034}.
  \href{http://arxiv.org/abs/1104.2933}{{\tt arXiv:1104.2933}}.
\bibitem[{Torrado and Lewis(2020)}]{Torrado:2020dgo}
\bibinfo{author}{J.~Torrado}, \bibinfo{author}{A.~Lewis},
\newblock \bibinfo{title}{{Cobaya: Code for Bayesian Analysis of hierarchical
  physical models}}  (\bibinfo{year}{2020}).
  \href{http://arxiv.org/abs/2005.05290}{{\tt arXiv:2005.05290}}.
\bibitem[{Lewis and Bridle(2002)}]{Lewis:2002ah}
\bibinfo{author}{A.~Lewis}, \bibinfo{author}{S.~Bridle},
\newblock \bibinfo{title}{{Cosmological parameters from CMB and other data: A
  Monte Carlo approach}},
\newblock \bibinfo{journal}{Phys. Rev. D} \bibinfo{volume}{66}
  (\bibinfo{year}{2002}) \bibinfo{pages}{103511}.
  \DOIprefix\doi{10.1103/PhysRevD.66.103511}.
  \href{http://arxiv.org/abs/astro-ph/0205436}{{\tt arXiv:astro-ph/0205436}}.
\bibitem[{Lewis(2013)}]{Lewis:2013hha}
\bibinfo{author}{A.~Lewis},
\newblock \bibinfo{title}{{Efficient sampling of fast and slow cosmological
  parameters}},
\newblock \bibinfo{journal}{Phys. Rev. D} \bibinfo{volume}{87}
  (\bibinfo{year}{2013}) \bibinfo{pages}{103529}.
  \DOIprefix\doi{10.1103/PhysRevD.87.103529}.
  \href{http://arxiv.org/abs/1304.4473}{{\tt arXiv:1304.4473}}.
\bibitem[{Handley et~al.(2015)Handley, Hobson, and Lasenby}]{Handley:2015fda}
\bibinfo{author}{W.~Handley}, \bibinfo{author}{M.~Hobson},
  \bibinfo{author}{A.~Lasenby},
\newblock \bibinfo{title}{{PolyChord: nested sampling for cosmology}},
\newblock \bibinfo{journal}{Mon. Not. Roy. Astron. Soc.} \bibinfo{volume}{450}
  (\bibinfo{year}{2015}) \bibinfo{pages}{L61--L65}.
  \DOIprefix\doi{10.1093/mnrasl/slv047}.
  \href{http://arxiv.org/abs/1502.01856}{{\tt arXiv:1502.01856}}.
\bibitem[{{Handley} et~al.(2015){Handley}, {Hobson}, and
  {Lasenby}}]{2015MNRAS.453.4384H}
\bibinfo{author}{W.~J. {Handley}}, \bibinfo{author}{M.~P. {Hobson}},
  \bibinfo{author}{A.~N. {Lasenby}},
\newblock \bibinfo{title}{{POLYCHORD: next-generation nested sampling}},
\newblock \bibinfo{journal}{Mon. Not. Roy. Astron. Soc.} \bibinfo{volume}{453}
  (\bibinfo{year}{2015}) \bibinfo{pages}{4384--4398}.
  \DOIprefix\doi{10.1093/mnras/stv1911}.
  \href{http://arxiv.org/abs/1506.00171}{{\tt arXiv:1506.00171}}.
\bibitem[{Foreman-Mackey et~al.(2013)Foreman-Mackey, Hogg, Lang, and
  Goodman}]{ForemanMackey:2012ig}
\bibinfo{author}{D.~Foreman-Mackey}, \bibinfo{author}{D.~W. Hogg},
  \bibinfo{author}{D.~Lang}, \bibinfo{author}{J.~Goodman},
\newblock \bibinfo{title}{{emcee: The MCMC Hammer}},
\newblock \bibinfo{journal}{Publ. Astron. Soc. Pac.} \bibinfo{volume}{125}
  (\bibinfo{year}{2013}) \bibinfo{pages}{306--312}.
  \DOIprefix\doi{10.1086/670067}. \href{http://arxiv.org/abs/1202.3665}{{\tt
  arXiv:1202.3665}}.
\bibitem[{Lewis(2019)}]{Lewis:2019xzd}
\bibinfo{author}{A.~Lewis},
\newblock \bibinfo{title}{{GetDist: a Python package for analysing Monte Carlo
  samples}}  (\bibinfo{year}{2019}).
  \href{http://arxiv.org/abs/1910.13970}{{\tt arXiv:1910.13970}}.
\bibitem[{Perenon et~al.(2019)Perenon, Bel, Maartens, and de~la
  Cruz-Dombriz}]{Perenon:2019dpc}
\bibinfo{author}{L.~Perenon}, \bibinfo{author}{J.~Bel},
  \bibinfo{author}{R.~Maartens}, \bibinfo{author}{A.~de~la Cruz-Dombriz},
\newblock \bibinfo{title}{{Optimising growth of structure constraints on
  modified gravity}},
\newblock \bibinfo{journal}{JCAP} \bibinfo{volume}{06} (\bibinfo{year}{2019})
  \bibinfo{pages}{020}. \DOIprefix\doi{10.1088/1475-7516/2019/06/020}.
  \href{http://arxiv.org/abs/1901.11063}{{\tt arXiv:1901.11063}}.
\bibitem[{Nesseris et~al.(2017)Nesseris, Pantazis, and
  Perivolaropoulos}]{Nesseris:2017vor}
\bibinfo{author}{S.~Nesseris}, \bibinfo{author}{G.~Pantazis},
  \bibinfo{author}{L.~Perivolaropoulos},
\newblock \bibinfo{title}{{Tension and constraints on modified gravity
  parametrizations of $G_{\textrm{eff}}(z)$ from growth rate and Planck data}},
\newblock \bibinfo{journal}{Phys. Rev. D} \bibinfo{volume}{96}
  (\bibinfo{year}{2017}) \bibinfo{pages}{023542}.
  \DOIprefix\doi{10.1103/PhysRevD.96.023542}.
  \href{http://arxiv.org/abs/1703.10538}{{\tt arXiv:1703.10538}}.

\end{thebibliography}
\end{document}